\documentclass[reqno]{amsart}
\usepackage{amsmath,amssymb,graphics,epsfig,color,cite}

\newtheorem{theorem}{Theorem}[section]   
\newtheorem{proposition}[theorem]{Proposition}
\newtheorem{lemma}[theorem]{Lemma}

\numberwithin{equation}{section}
\numberwithin{theorem}{section}

\renewcommand{\hat}{\widehat}

   \newcommand{\nn}{\nonumber}

\newcommand{\mc}[1]{{\mathcal #1}}

\newcommand{\bb}[1]{{\mathbb #1}}

\newcommand{\ol}[1]{\,\overline {\!#1\!}\,}

\renewcommand{\epsilon}{\varepsilon}
\newcommand{\id}{{1 \mskip -5mu {\rm I}}}
 
\newcommand{\varsh}{\mathop{\rm sh}\nolimits}
\newcommand{\varch}{\mathop{\rm ch}\nolimits}
\newcommand{\varth}{\mathop{\rm th}\nolimits}
\newcommand{\avarsh}{\mathop{\rm arcsh}\nolimits}
\newcommand{\avarth}{\mathop{\rm arcth}\nolimits}
\newcommand{\sgn}{\mathop{\rm sgn}\nolimits}

\newcommand{\bam}{\,\overline {\!m\!}\,}

\begin{document}

\title
{Dobrushin states in the $\boldsymbol{\phi^4_1}$ model}

\author [L.\ Bertini] {Lorenzo Bertini}
\address{Lorenzo Bertini, Dipartimento di Matematica,
Universit\`a di Roma `La Sapienza', P.le Aldo Moro 2,
00185 Roma, Italy}
\email{bertini@mat.uniroma1.it}

\author [S.\ Brassesco] {Stella Brassesco}
\address{Stella Brassesco, Departamento de Matem\'aticas, 
Instituto Venezolano de Investigaciones Cient\'{\i}ficas,
Apartado Postal 21827, Caracas 1020--A, Venezuela}
\email{sbrasses@ivic.ve}

\author [P.\ Butt\`a] {Paolo Butt\`a}
\address{Paolo Butt\`a, Dipartimento di Matematica,
Universit\`a di Roma `La Sapienza', P.le Aldo Moro 2,
00185 Roma, Italy}
\email{butta@mat.uniroma1.it}

%\thanks{}

\begin{abstract}
We consider the van der Waals free energy functional in a bounded
interval with inhomogeneous Dirichlet boundary conditions imposing
the two stable phases at the endpoints. We compute the asymptotic
free energy cost, as the length of the interval diverges, of
shifting the interface from the midpoint. 
We then discuss the effect of thermal fluctuations by analyzing
the $\phi^4_1$-measure with Dobrushin boundary conditions. In
particular, we obtain a nontrivial limit in a suitable scaling
in which the length of the interval diverges and the temperature vanishes. The limiting state is not translation invariant
and describes a localized interface. This result can be seen as
the probabilistic counterpart of the variational convergence of 
the associated excess free energy.  
\end{abstract}


\maketitle

\section{Introduction}
\label{sec:1}
\thispagestyle{empty}

The van der Waals' theory of phase transition is based on the
functional
   \begin{equation}
   \label{acf}
\mathcal{F}(m) = \int\!dx\, \Big[ \frac 12 m'(x)^2
+ 2 V(m(x)) \Big],
   \end{equation}
where the scalar field $m(x)$ represents the local order parameter
and $V(m)$ is a smooth, symmetric, double well potential whose 
minimum value, chosen to be zero, is attained at $m_\pm$; we also
assume $V''(m_\pm)>0$. We restrict the discussion
to the one dimensional case $x\in\bb R$. If \eqref{acf} is considered 
in the whole line $\bb R$, there are infinitely many critical points. 
The most relevant ones are the constant profiles $m_\pm$, where 
$\mathcal{F}$ attains its minimum, and $\pm\bam(x)$, where $\bam(x)$
is the  solution to  
   \begin{equation}
   \label{1.1bis}
\bam''(x)- 2 V'(\bam(x)) =0, \qquad \lim_{x\to
\pm\infty} \bam(x) = m_\pm, \qquad \bam(0) = 0,
   \end{equation}
together with its translates $\pm\bam_z(x) =\pm \bam(x-z)$,
$z\in\bb R$. Note that $\bam_z$ minimizes $\mathcal{F}$ under 
the constraint that $\lim_{x\to \pm\infty} m(x) = m_\pm$. Therefore 
$\bam_z$ is the stationary profile with the two pure phases 
$m_\pm$ coexisting to the right and to the left of $z$. Accordingly,
the van der Waals surface tension is $\sigma_w = \mc F(\bam)$.
We set $\mathcal{M} = \{\bam_z : z\in {\bb R}\}$. We
emphasize that we do not consider the sharp interface limit which is
obtained by  introducing a scaling parameter in \eqref{acf}. In particular,
even if the convergence of $\bam_z$ to its asymptotic values is 
exponentially fast, the profile $\bam_z$ describing the interface 
is not sharp but diffuse, we refer to it as a \emph{mesoscopic} 
interface. 

Our first purpose is to analyze the finite size effects in the free 
energy $\mathcal{F}$. More precisely, we consider \eqref{acf} in
the bounded interval $[-\ell,\ell]$ with the inhomogeneous Dirichlet 
boundary conditions $m(\pm\ell) = m_\pm$. If we think of $m$ as the local magnetization, this condition models the effect of opposite magnetic fields applied at the endpoints. We denote by 
$\mathcal{F}_\ell$ the functional \eqref{acf} with these stipulated 
boundary conditions.

It is quite easy to show that the functional $\mc F_\ell$ has a 
unique minimizer $m^*_\ell$, which by symmetry converges to $\bam_0$ as
$\ell\to\infty$. On the other hand, the limiting functional 
$\mathcal{F}$ is minimized, under the constraint $m(\pm\infty)=m_\pm$,
by any shifted interface $\bam_z\in\mc M$. It is therefore natural to
introduce the \emph{excess} free energy 
   \begin{equation}
   \label{gll0}
\mc G_\ell(m) =
e^{\alpha\ell} \big[ \mc F_\ell(m) - \mc F_\ell(m^*_\ell)\big],  
   \end{equation}
in which the exponential rescaling $e^{\alpha\ell}$ is chosen to get a
nontrivial limit as $\ell\to\infty$. Indeed, in this paper we show
there exists $\alpha=\alpha(V)$ for which $\mc G_\ell$ converges
to a limiting functional $\mc G$ which is finite only on the set 
$\mc M$, where it is given by 
$$
\mc G(\bam_z) =  A \big[ \varch(\alpha z) -1\big],
$$
for a suitable constant $A=A(V)>0$. 
The quantity $e^{-\alpha\ell} \mc G(\bam_z)$ gives therefore the
asymptotic free energy cost needed to shift the interface by $z$ 
and encodes the leading finite size correction to the free energy 
$\mc F$. 

\medskip
Actually, the above variational formulation of phase transitions
neglects completely the microscopic fluctuations, which play an 
important role in  various phenomena. At the mesoscopic level,
the effect of fluctuations can be modeled by considering the 
probability measure, on the space of order parameter profiles, 
informally given by
\begin{equation}
\label{5.2.1}
d\mu_{\epsilon}(m) 
= Z^{-1}\, \exp\big\{-\epsilon^{-1}\mathcal{F}(m)\big\}
\, \prod_x dm(x).
\end{equation}
In the case $V(m) = \frac 14(m^2-1)^2$, the above measure corresponds 
to the Euclidean version of the quantum anharmonic oscillator and it is
usually referred to as the $\phi_1^4$-measure, here the subscript one 
stands for one dimension. This model has been extensively 
analyzed because exhibits an interesting
behavior in a simple setting, see \cite{Simon} and references therein. 

In the van der Waals theory, the local order parameter $m(x)$ represents
the empirical average, on a mesoscopic scale, of the microscopic observable.
Accordingly, the parameter $\epsilon$ is to be interpreted as the ratio
between the microscopic scale (say of the order of Angstroms) to the 
mesoscopic one (say of the order of tens of microns). In this Gibbsian setting, the chosen inhomogeneous Dirichlet boundary conditions 
are usually referred to as Dobrushin boundary conditions, their effect 
is to force an interface in the system. We denote by $\mu_{\epsilon,\ell}$
the probability measure defined as in \eqref{5.2.1} with $\mc F$
replaced by $\mc F_\ell$. For $\ell$ fixed and $\epsilon$ small, 
since the measure $\mu_{\epsilon,\ell}$ concentrates on the minimizers 
of $\mathcal{F}_\ell$, a typical configuration is close to $m^*_\ell$.
On the other hand, since the model is one dimensional, 
for $\epsilon$ fixed and $\ell\to\infty$ the measure 
$\mu_{\epsilon,\ell}$ forgets the prescribed boundary conditions and
converges to the unique infinite  volume Gibbs state.
The precise statement would be that the measure $\mu_{\epsilon,\ell}$,
considered on $C(\bb R)$ with the topology of uniform convergence in
compacts, converges weakly as $\ell\to\infty$ to an infinite
volume Gibbs measure, defined as a solution to the DLR equations.
In the context of the $\phi_4^1$ model, uniqueness of solution to 
the DLR equations follows from the analysis in \cite[II.6]{Simon}, 
but we did not find in the literature a detailed proof (see however
the discussion in \cite[\S~II.5,~VII.2]{GRS}) of the 
weak convergence of $\mu_{\epsilon,\ell}$ to the unique infinite
volume state. However this is not really relevant in
the present paper, in which we investigate a diagonal limit 
$\epsilon\to 0$ and $\ell\to \infty$. In particular, the aforementioned 
convergence of the excess free energy $\mc G_\ell$ suggests that 
a nontrivial limiting behavior could be obtained by choosing 
$\epsilon = e^{-\alpha\ell}$.
We show this is indeed the case: with this choice, the measure
$\mu_{\epsilon,\ell}$ weakly converges to a measure $\mu$ with support
$\mc M$ and there given by 
\begin{equation}
\label{5.2.2}
d\mu(\bam_z) = N^{-1}\, \exp\{-\mc G(\bam_z)\} \, dz.
\end{equation}
We call this limiting measure a Dobrushin state because it is not
translation invariant and describes a fluctuating interface. We 
emphasize however that the order parameter profile is fixed and, 
with probability super-exponentially close to one as $L\to \infty$, 
the interface is localized in the bounded interval $(-L,L)$.

In the case of short range, ferromagnetic, lattice models of 
statistical mechanics (Ising models), phase transitions may occur
only in dimension $d\ge 2$. The behavior of interface 
fluctuations when the system is considered in a box of side $\ell$ 
and Dobrushin boundary conditions are imposed has been analyzed in 
detail, see e.g.\ \cite{Presutti} for a review. In $d=2$
the interface behaves as a random walk having fluctuations of the
order of $\sqrt\ell$; in particular, in the thermodynamic limit
$\ell\to\infty$, the corresponding Gibbs measure converges to a
translation invariant state which is a mixture of the pure phases,
i.e.\ there are no Dobrushin states \cite{DS}. In $d\ge 3$, for low
temperature, the interface fluctuations remain bounded and a not translation invariant state is obtained in the thermodynamic limit
\cite{D}. With respect to the above
context, the diagonal limit $\ell\to\infty$, $\epsilon\to 0$, 
corresponds to a joint limit in which the size of the system diverges
and the temperature vanishes. This peculiar limiting procedure
allows to get nontrivial Dobrushin states for $d<3$. We also mention
that a localized interface can be obtained for long-range (power
law decay) one-dimensional Ising models \cite{CMR}. 

   \section{Notation and results}
   \label{sec:2}
It will be convenient to denote by $t$ the space variable
and by $x=x(t)$ a continuous function of $t$. Let
   \begin{equation}
   \label{mX=}
\mc X := \big\{x\in C(\bb R) \, : \, \lim_{t\to\pm\infty} x(t) = \pm 1 \big\},
   \end{equation} 
endowed with the metric $d(x,y) := \|x-y\|_\infty :=
\sup_{t\in \bb R} |x(t)-y(t)|$ and the associated Borel 
$\sigma$-algebra. We emphasize that we need to use this topology 
and not the one of uniform convergence on compacts because we need 
to distinguish the behavior of $x$ as $|t| \to \infty$. 
Given $\ell>0$, we also let 
   $$
\mc X_\ell := \Big\{x\in C(\bb R) \, : \: x(t) = \sgn(t) \quad
\text{for } |t|\ge \ell \Big\},
   $$ 
which is a closed subset of $\mc X$. 
 
For the sake of concreteness, in this paper we restrict the analysis 
to the paradigmatic case of the symmetric double well potential, 
i.e.\ we choose
    \begin{equation}
    \label{p1}
V(x) = \frac 14 \big(x^2-1\big)^2,
    \end{equation}
which attains its minimum at $x=\pm 1$. In this case the solution
to \eqref{1.1bis} is given by $\bam(t) = \varth(t)$; for $z\in\bb R$ 
we set $\bam_z(t) := \varth(t-z)$, and define 	
\begin{equation}
	\label{mcM=}
\mc M:= \{\bam_z : z \in {\bb R}\},
	\end{equation}  
which is a closed subset of $\mc X$.
Given $\ell>0$, we denote by $W^{1,2}_\ell$ the Sobolev space
$W^{1,2}([-\ell,\ell])$ and define the finite volume free energy as
the functional $\mc F_\ell : \mc X \to [0,+\infty]$ given by
   \begin{equation}
   \label{fell}
\mc F_\ell(x) := 
\int_{-\ell}^{+\ell}\!dt\, \Big[ \frac 12\, x'(t)^2 + 2V(x(t)) \Big]
   \end{equation}
if $x\in\mc X_\ell$ and $x\!\restriction_\ell \, \in W^{1,2}_\ell$,
while $\mc F_\ell(x) := +\infty$ otherwise. Here 
$x\!\restriction_\ell$ denotes the restriction of $x$ to 
$(-\ell,\ell)$. 

Our first statement concerns the limiting behavior
of the sequence $\mc F_\ell$. This result can be seen as a {\it
diffuse} version of the classical Modica-Mortola result, see e.g.\
\cite[Thm.\ 6.4]{Braides}. More precisely, the latter result deals
with the sharp interface limit, and states that the limiting 
free energy is concentrated on profiles taking values in
$\{-1;1\}$ and counts the number of jumps. Here we instead show that
any minimizer of the limiting functional $\mc F$ is a profile in 
$\mc M$.

Referring e.g.\ to \cite[Ch.~1]{Braides} for more details, we next
outline the basic definitions and results of the $\Gamma$-convergence
theory. Let $X$ be a metric space.  A sequence of functionals $F_n : X
\to [0,+\infty]$ is \emph{equi-coercive} iff from any sequence $x_n$
such that $\varlimsup_n F_n(x_n) < +\infty$ it is possible to extract
a converging subsequence. The sequence $F_n$ is \emph{equi-mildly
coercive} iff there exists an non-empty compact set $K\subset X$
such that $\inf_X F_n = \inf_KF_n$ for any $n\in\bb N$. The sequence
$F_n$ $\Gamma$-converges to a functional $F : X \to [0,+\infty]$ iff
the following conditions hold for each $x\in X$. There exists a
sequence $x_n\to x$ such that $\varlimsup_n F_n(x_n) \le F(x)$
(\emph{$\Gamma$-limsup inequality}) and for any sequence $x_n\to x$ we
have $\varliminf_n F_n(x_n) \ge F(x)$ (\emph{$\Gamma$-liminf
  inequality}).  If the sequence $F_n$ is mildly equi-coercive and
$\Gamma$-converges to $F$ then $\inf_XF = \min_XF = \lim_n \inf_X
F_n$. Moreover, if $x_n$ is a pre-compact sequence such that $\lim_n
F_n(x_n) = \lim_n \inf_XF_n$ then every converging subsequence of $x_n$
is a minimizer of $F$. Finally, if the sequence $F_n$ is equi-coercive
and $\Gamma$-converges to $F$ then, for each open set $A$ and each
closed set $C$ we have
$$
\varlimsup_n \; \inf_A F_n \; \le \; \inf_A F, \quad\qquad
\varliminf_n \; \inf_C F_n \; \ge \; \inf_C F,
$$
which are the relevant estimates in the asymptotic analysis of the
free energy. 

  \begin{theorem}
  \label{gellconv}
The sequence $\mc F_\ell : \mc X \to [0,+\infty]$ is equi-mildly coercive and as $\ell\to\infty$ $\Gamma$-converges to 
   $$
\mc F(x) := \begin{cases} {\displaystyle \int_{-\infty}^{+\infty}\!dt\, 
\Big[ \frac 12\, x'(t)^2 + 2V(x(t)) \Big]} & \text{if $x' \in L_2(\bb R)$ and 
$1-x^2 \in L_2(\bb R)$,} \\ +\infty & \text{otherwise.} \end{cases}
   $$
Moreover, the set of minimizers of $\mc F$ is $\mc M$, as defined in 
\eqref{mcM=}. In particular, the (van der Waals) surface tension
is
\begin{equation}
\label{sigma}
\sigma_w = \inf_{\mc X} \mc F = \mc F(\bam) = \frac 43. 
\end{equation}
   \end{theorem}

We remark that $\mc F_\ell$ is not equi-coercive. Indeed, we can
construct a diverging numeric sequence $z_\ell$ and a sequence
$x_\ell$ such that $\|x_\ell-\bam_{z_\ell}\|_\infty \to 0$ and
$\mc F_\ell (x_\ell)\to 0$. 
As stated in the previous theorem, the limiting free energy $\mc F$
does not remember that $m^*_\ell$, the unique minimizer of $\mc
F_\ell$, converges to $\bam_0$. The underlying reason is that the
finite volume free energy cost of profiles close to $\bam_z$,
$z\in\bb R$, is infinitesimal as $\ell\to\infty$. We then introduce 
the \emph{excess free energy} $\mc G_\ell : \mc X \to [0,+\infty]$
as
   \begin{equation}
   \label{gll}
\mc G_\ell(x) := e^{4\ell}\,\big[ \mc F_\ell(x) - 
\mc F_\ell(m^*_\ell)\big],
   \end{equation}
in which the rescaling $e^{4\ell}$ has been chosen to get a nontrivial
limit as $\ell\to\infty$. In fact, as shown in 
Proposition~\ref{t:41} below, the finite volume corrections
to the surface tension are $O(e^{-4\ell})$, in particular
$\lim_\ell e^{4\ell}\big[ \mc F_\ell(m^*_\ell) - \sigma_w  \big] = 16$. 
In this setting the limiting
functional $\mc G$ will be finite only on $\mc M$ and describes
the asymptotic cost of shifting an interface from the origin.
Indeed, in the next theorem we identify the $\Gamma$-limit of 
$\mc G_\ell$. This is usually referred to as the {\em development 
by $\Gamma$-convergence}.

   \begin{theorem}
   \label{barfell}
The sequence $\mc G_\ell : \mc X \to [0,+\infty]$ is equi-coercive 
and as $\ell\to\infty$ $\Gamma$-converges to 
   \begin{equation}
   \label{glim}
\mc G(x) := \begin{cases} 16 \big[ \varch(4z) -1\big] & \text{ if 
$x = \bam_{z}$ for some $z \in\bb R$,} \\ 
+ \infty & \text{ otherwise.} \end{cases}
   \end{equation}
   \end{theorem}
    
\medskip
We now discuss the asymptotic behavior of the $\phi_1^4$-measure 
with Dobrushin boundary conditions. We first recall the precise
definition of the measure informally introduced in \eqref{5.2.1}.
Given $\epsilon >0$ we denote by $\varrho_{\epsilon,\ell}$ the 
probability measure on $\mc X$, whose support
is $\mc X_\ell$ and having there  the law of the Brownian bridge
with diffusion coefficient $\epsilon$, starting at time $-\ell$ 
from $-1$ and arriving at time $\ell$ to $+1$. In other words, 
$\varrho_{\epsilon,\ell}$ is the Gaussian measure on $\mc X$ with 
mean 
   $$
\ol x_\ell(t) := \varrho_{\epsilon,\ell}\big(x(t)\big) 
= \begin{cases} \frac t\ell & \text{ if } |t|\le\ell, \\
\sgn(t) & \text{ if } |t|>\ell, \end{cases} 
   $$
and covariance
   $$
\varrho_{\epsilon,\ell}\Big( \big[x(t)-\ol x_\ell(t)\big]\,
\big[x(s)-\ol x_\ell(s)\big]\Big) = 
\begin{cases} \frac{\epsilon}{2\ell} \, 
(\ell+s\wedge t)(\ell-s\vee t) & 
\text{ if } s,t\in [-\ell,\ell], \\
0 & \text{ otherwise,} \end{cases} 
   $$
where hereafter $\mu(f)$ denotes the expectation of the measurable 
function $f$ w.r.t.\ the measure $\mu$ and, for $a,b\in\bb R$, 
$a\land b$ (resp.\ $a\vee b$) denotes the minimum (resp.\ maximum) 
between $a$ and $b$.

The $\phi^4_1$ model at temperature
$\epsilon$ with Dobrushin type boundary condition is the probability measure 
$\mu_{\epsilon,\ell}$ on $\mc X$ with support $\mc X_\ell$, whose 
density w.r.t.\ 
$\varrho_{\epsilon,\ell}$ is given by
\begin{equation}
\label{p2}
\frac{d\mu_{\epsilon,\ell}}{d\varrho_{\epsilon,\ell}} (x) =
\frac{1}{Z_{\epsilon,\ell}} \, \exp\Big\{-\epsilon^{-1} 
\int_{-\ell}^\ell\!dt\, 2V(x(t)) \Big\},
\end{equation}
where 
\begin{equation}
\label{Z=}
Z_{\epsilon,\ell} := 
\varrho_{\epsilon,\ell}\Big(\exp\Big\{-\epsilon^{-1} 
\int_{-\ell}^\ell\!dt\, 2V(x(t)) \Big\}\Big).
\end{equation}
From the Laplace-Varadhan theorem it follows, see e.g.\ \cite[Ex.~4.3.11]{DZ}, that
for $\ell$ fixed the probability $\mu_{\epsilon,\ell}$ satisfies
a large deviation principle with speed $\epsilon^{-1}$ and rate function
$\mc F_\ell(x) - \mc F_\ell(m^*_\ell)$. On the other hand, by
Theorem~\ref{barfell}, the functional  
$\mc F_\ell(x) - \mc F_\ell(m^*_\ell)$ behaves like $e^{-4\ell}
\mc G(x)$. Therefore we expect that, in the diagonal limit
$\ell = \frac 14 \log\epsilon^{-1}$ and $\epsilon \to 0$, the measure 
$\mu_{\epsilon,\ell}$ converges to a non-degenerate limit $\mu$,
which should look like $d\mu(x) \approx e^{-\mc G(x)}dx$. 
Our main result shows that this is indeed the case.

\begin{theorem}
\label{t:1}
Let $\ell = \frac 14 \log\epsilon^{-1}$, then the measure
$\mu_{\epsilon,\ell}$ converges weakly in $\mc X$ as $\epsilon\to 0$ 
to the measure $\mu$ defined on the  Borel sets $A\subset \mc X$ by 
\begin{equation}
  \label{mecm}
  \mu(A) = \hat\mu\big(\big\{z\in\bb R \, : \, \bam_z\in A\big\}\big),
\end{equation}
where $\hat\mu$ is the probability measure on $\bb R$ given by
  \begin{equation}
  \label{hatmu}  
\hat\mu(dz) =  \frac{e^{- 16 [\varch(4z)-1]}}
{\int\!dz'\, e^{- 16 [\varch(4z')-1]}}\: dz.
 \end{equation}
  \end{theorem}

We emphasize that this nontrivial limiting behavior is due to the
particular choice $\ell= \frac 14 \log\epsilon^{-1}$, the
coefficient $\frac 14$  coming from the specific form \eqref{p1} of the
double well potential $V$. From the analysis carried out in this paper
it follows that if we had considered $\ell= \big(\frac 14-\delta\big)
\log\epsilon^{-1}$ for some $\delta>0$, the measure
$\mu_{\epsilon,\ell}$ would have converged weakly to the probability
concentrated on the single configuration $\bam_0$. Moreover,
it should be also possible to show that if we had  considered 
$\ell=\big(\frac 14+\delta\big)\log\epsilon^{-1}$ for some $\delta>0$, 
then the family $\mu_{\epsilon,\ell}$ would not have been tight on 
$\mc X$. On the other hand, the family $\mu_{\epsilon,\ell}$, still for $\ell=\big(\frac 14+\delta\big)\log\epsilon^{-1}$ and considered in  
$C(\bb R)$ endowed with the topology of uniform convergence on compacts, 
would converge  weakly to $\frac 12 [\delta_{-1}+\delta_{1}]$, where 
$\delta_{\pm 1}$ denotes the Dirac measure concentrated on the 
configuration identically equal to $\pm 1$: on this scale the 
interface ``went to infinity''. 
As it appears clear from the above discussion, the compactness property
of the family $\mu_{\epsilon,\ell}$ is a key point; in particular
tightness of $\mu_{\epsilon,\ell}$ implies that the interface remains
localized in compact subsets of $\bb R$.  

\smallskip
\noindent\emph{Strategy of the proof.}
As it is  well known, see e.g.\ \cite{Simon}, the measure describing the 
$\phi^4_1$ model in the whole line can be realized as the law of the
stationary process associated to the one-dimensional stochastic
differential equation 
\begin{equation}
  \label{eq:1}
  dX_t = a_\epsilon(X_t)dt + \sqrt\epsilon \, dw_t
\end{equation}
where $w_t$ is a standard Brownian motion and the drift $a_\epsilon$
is the logarithmic derivative of the ground state of the (quantum)
anharmonic oscillator. More precisely, let us denote by
$\lambda_\epsilon$ the smallest eigenvalue of the Schroedinger
operator $H_\epsilon:= -\frac 12\epsilon^2\Delta + 2V$ on 
$L_2(\bb R,dx)$,
the corresponding eigenfunction, chosen strictly positive, is denoted
by $\phi_\epsilon$. Then $a_\epsilon = \epsilon \nabla \log
\phi_\epsilon$; in particular $\phi_\epsilon(x)^2 dx$ is the invariant
measure of the process $X_t$. We mention that this representation
of the infinite volume $\phi_1^4$-measure allows, by means of 
Friedlin-Wentzell large deviations estimates \cite{FV}, a detailed study
of the typical configurations as $\epsilon\to 0$. From the analysis in
\cite{JMS}, whose main motivation lies on semiclassical limits,
the following picture emerges. With probability exponentially
close to one as $\epsilon \to 0$, we see $x(t) \approx \pm 1$
for $t$ in intervals of the order $e^{\epsilon^{-1}\sigma_w}$; the 
transition (tunneling) between the pure phases taking place 
in a small neighborhood of $\bam_z$ for suitable $z$'s. Moreover,
if the lengths of the above intervals are properly normalized, 
they converge weakly to an independent jump process with 
exponential distribution, as in the case of Ising spin systems, 
either nearest-neighbors \cite{ST} or with long range interaction
of Kac type \cite{COP}. 

A representation in terms of a diffusion process can be obtained also 
in the present setting of the $\phi^4_1$ model with
Dobrushin boundary condition. From a statistical mechanics viewpoint,
this representation corresponds to transfer matrix arguments. The
probability $\mu_{\epsilon,\ell}$ can be realized as
the law of the diffusion process \eqref{eq:1} with initial condition
$X_{-\ell}=-1$ \emph{conditioned} to reach $1$ at the time
$t=\ell$. According to the results in \cite{HP}, this conditioned
process can be also realized as the solution to a stochastic
differential equation with a time dependent drift. Let us denote by
$X^{x}_t$ the solution to \eqref{eq:1} with initial condition
$X_{0}=x$ and introduce the transition probability density
$p^\epsilon_{t}(x,y)$ by requiring that for each $t>0$ and each
Borel set $B\subset \bb R$,
\begin{equation}
  \label{eq:2.0}
P \big(X^{x}_t \in B \big) = \int_B\!dy\, p^\epsilon_{t}(x,y).
\end{equation}
For $(t,x)\in (-\ell,\ell)\times \bb R$, we define
\begin{equation}
  \label{eq:2.1*}
\hat a_{\epsilon,\ell} (t,x) := -a_\epsilon(x) 
+ \epsilon \partial_x \log p^\epsilon_{\ell-t}(1,x).
\end{equation}
Then, as follows from \cite{HP}, the measure $\mu_{\epsilon,\ell}$ is
the law of the process $Y$ defined as follows.  For $|t|\ge \ell$ we
set $Y_t=\sgn(t)$ while for $|t|< \ell$ we define $Y_t$ as the
solution to the stochastic differential equation 
\begin{equation}
  \label{eq:2}
  \begin{cases}
    dY_t = \hat a_{\epsilon,\ell} (t,Y_t) dt +\sqrt{\epsilon}\, dw_t, \\
    Y_{-\ell}=-1,
  \end{cases}  
\end{equation}
here $w_t$, $t\in [-\ell,\ell]$, is a standard Brownian with $w_{-\ell}=0$.

Theorem~\ref{t:1} can therefore, equivalently, be rephrased in terms
of the limiting behavior of the solution to \eqref{eq:2}. We emphasize
that we obtain a non-degenerate limiting behavior as $\epsilon\to 0$
even if the noise term vanishes. This is due both to the simultaneous
divergence of the time interval and to the peculiar behavior of the
drift $\hat a_{\epsilon,\ell}$. We discuss the latter issue in some
more detail. By the well known ground state transformation, see e.g.\
\cite{Simon}, we can rewrite the transition probability density in
\eqref{eq:2.0} in terms of the kernel of the semigroup generated by
$H_\epsilon$,
\begin{equation*}
p^\epsilon_t(x,y) = \frac{1}{\phi_\epsilon(x)}\,
e^{-\epsilon^{-1}(H_\epsilon -\lambda_\epsilon)t}(x,y) \, \phi_\epsilon(y),
\end{equation*}
so that, recalling \eqref{eq:2.1*}, 
\begin{equation}
\label{222}
\hat a_{\epsilon,\ell}(t,x) = \epsilon \partial_x \log
e^{-\epsilon^{-1}(H_\epsilon -\lambda_\epsilon)(\ell-t)}(1,x).
\end{equation}
It is also not difficult to check that, by writing 
$\hat a_{\epsilon,\ell}(t,x) = - \partial_xS_\epsilon(\ell-t,x)$, 
the function $S_\epsilon$ solves the viscous Hamilton-Jacobi
equation
\begin{equation}
\label{se}
\partial_t S_\epsilon + \frac 12 (\partial_xS_\epsilon)^2-2V
= \frac\epsilon 2 \Big[ \partial_{xx}S_\epsilon - \frac 1t\Big],
\end{equation}
of course $S_\epsilon$ is singular as $t\downarrow 0$.  
To analyze the solution to \eqref{eq:2}, as $\epsilon\to 0$, we
therefore need sharp estimates on the semiclassical limit of
the Schrodinger operator $H_\epsilon$. More precisely, we need
good control on the kernel of the corresponding semigroup up to
times of order $\ell = O(\log\epsilon^{-1})$. In the context of semiclassical limits, see e.g.\ \cite{Robert}, this scale of time 
is known as \emph{Erhenfest time} and it is the one in which the semiclassical approximation is not - in general - anymore valid. 

As it appears quite intricate to get good control on 
$\hat a_{\epsilon,\ell}$
by direct semiclassical methods or perturbation theory in 
Hamilton-Jacobi, we follow a different approach, which we might call
Euclidean semiclassical approximation. 
If $\ell$ were fixed, by the Feynmann-Kac formula and Laplace-Varadhan
asymptotic in \eqref{222}, we would get, as $\epsilon\to 0$, 
\begin{equation}
\label{alim}
\hat a_{\epsilon,\ell}(t,x) \approx - \partial_x S(\ell-t,x),
\end{equation} 
where   
	\begin{equation}
   \label{S=}
S(t,x) := \inf\Big\{ \int_0^t\!ds\, \Big[
\frac 12 \dot\psi(s)^2 + 2V(\psi(s))\Big]
\, : \: \psi(0) = 1, \, \psi(t) = x \Big\}
   	\end{equation}
is the action for a Newtonian particle of mass one in the
potential $-2V$ starting at time zero from $1$ and arriving
at time $t$ to $x$; the change of sign in the potential is due 
to the fact that we are looking at the Schroedinger semigroup.
Note that $S$ solves \eqref{se} with $\epsilon=0$.
As the r.h.s.\ of \eqref{alim} makes sense, we use it
as the drift term of an auxiliary diffusion process. Namely, 
we introduce the process $\xi$ as the solution to
\begin{equation*}
d\xi_t =  - \partial_x S(\ell-t,\xi_t) \, dt + \sqrt\epsilon\, dw_t,
\end{equation*}
where $w$ is a standard Brownian motion. 
Since \eqref{alim} is not an identity, the law of $\xi$ is not
$\mu_{\epsilon,\ell}$. On the other hand it is a good approximation of
it in the sense that, as shown in Proposition~\ref{t:61} below, 
their Radon-Nykodim derivative is ``only'' of the order $e^{O(\ell)}$. 
Moreover, even if the drift term above is not really given explicitly, 
standard methods for one dimensional mechanical systems allow to get
sharp estimates on it. 

By exploiting the above strategy, we get enough
control on the measure $\mu_{\epsilon,\ell}$ to show that it
concentrates in a small neighborhood of $\mc M$ and that it is
tight in $\mc X$. The identification of its limit points with the
measure $\mu$ defined in Theorem~\ref{t:1} will be 
accomplished by a dynamical argument. We refer to
\cite{Funaki} for a recent review on the dynamics
of stochastic interfaces. The probability $\mu_{\epsilon,\ell}$ can
be in fact characterized, see \cite[Thm.\ 5.1]{FV}, as the unique
invariant measure of the Markov process $X\equiv 
X_\sigma(t)$, $(\sigma,t) \in \bb R_+\times\bb R$, 
in $C(\bb R_+;\mc X_\ell)$ which solves the stochastic partial
differential equation 
\begin{equation}
\label{feq-1}
\begin{cases}
d X_\sigma =\big[ \frac12\partial_{tt}\,X_\sigma-V'(X_\sigma) 
\big]d\sigma +\sqrt\epsilon\, dW_\sigma & \sigma >0, \, |t|<\ell, \\
X_\sigma(t)=\sgn(t) & \sigma\ge 0, \, |t|\ge \ell, \end{cases}
\end{equation}
where $W$ is the cylindrical Wiener process on 
$L_2([-\ell,\ell],dt)$.  
As shown in \cite{bbbdy}, in the scaling limit
$\ell=\frac 14\log\epsilon^{-1}$ and $\epsilon\to 0$, 
$X_{\epsilon^{-1}\sigma}$ converges in law to $\bam_{\zeta_\sigma}$
where $\zeta$ solves 
\begin{equation}
\label{feq-2}
d\zeta_\sigma = - \, 24\, \varsh\big(4\zeta_\sigma\big) d\sigma 
+ \sqrt{\frac 34} \, d B_\sigma,
\end{equation}
with $B$ a standard Brownian motion. As the unique 
invariant measure of this one dimensional diffusion process is 
$\hat\mu$, see \eqref{hatmu}, we conclude the identification. 

A final remark on the relationship between the equilibrium 
asymptotic stated in Theorem~\ref{t:1} and the above dynamical
result is due. A basic paradigm in non-equilibrium 
statistical mechanics is the Einstein relation which connects
dynamical transport coefficients and thermodynamic potentials, 
see e.g.\ \cite[I.8.8]{S}. The general
structure of this relation is [drift] $=$ $\frac 12$ $\cdot$
[diffusion] $\cdot$ [thermodynamic force], where the thermodynamic
force is minus the derivative of the free energy. It is worth 
noticing that such a relationship is verified also in the present
setting of a drift induced by the boundary conditions, namely
$$
- \, 24 \varsh(4z) = -\frac 12\cdot \frac 34 \cdot
\frac{d}{dz} 16\big[\varch(4z)-1\big].
$$

   \section{Asymptotic analysis of the free energy}
   \label{sec:4}
   
In this section we analyze the asymptotic behavior of the free energy
$\mc F_\ell$ and prove Theorems \ref{gellconv} and \ref{barfell}.
We start by showing that for each $\ell>0$ there exists a unique
minimizer $m^*_\ell$ of $\mc F_\ell$ and discuss its behavior as
$\ell\to\infty$. Recall that for the choice \eqref{p1} of $V$ we
have $\mc F(\bam) = \frac 43$.

\begin{proposition}
\label{t:41}
The functional $\mc F_\ell$ has a unique minimizer $m^*_\ell$ in 
$\mc X_\ell$. Moreover $m^*_\ell\restriction_\ell\in C^2((-\ell,\ell))$
and for $|t|\le \ell$ is the unique solution to the boundary value
problem
   \begin{equation}
   \label{ml}
\begin{cases} \frac 12 x''(t) - V'(x(t)) = 0, & t\in (-\ell,\ell), \\
x(\pm\ell)=\pm 1. & \end{cases}
   \end{equation}
Finally,
   \begin{eqnarray}
   \label{stima*}
&& \varlimsup_{\ell\to \infty}\, e^{2\ell} \, 
\big\|m^*_\ell - \bam_0 \big\|_\infty < \infty, \\ \label{stima**}
&& \lim_{\ell\to\infty} e^{4\ell} 
\Big[\mc F_\ell(m^*_\ell) - \frac 43 \Big] = 16.
   \end{eqnarray}
   \end{proposition}

\proof The boundary value problem \eqref{ml} is the Euler-Lagrange
equation for the stated variational problem. Equation \eqref{ml}
can be regarded as that of the motion of a Newtonian particle of
mass one in the potential $-2V$. 
By standard Weierstrass analysis of one-dimensional mechanical
systems it is then straightforward to prove that there exists a unique
twice differentiable solution $m^*_\ell$ to \eqref{ml}. Explicit 
estimates yield the bounds \eqref{stima*} and \eqref{stima**}, see
Lemma~\ref{t:33}. We here prove uniqueness of the minimizer with the given boundary conditions. The
argument is rather standard and it is reported for completeness. Let
us denote by $S_\sigma(x)$, $\sigma\ge 0$, the gradient flow associated 
to $\mc F_\ell$, i.e.\ $u(\sigma,t)=S_\sigma(x)(t)$ solves
   $$
\begin{cases}
{\displaystyle \partial_\sigma u(\sigma,t) = \frac12
\partial_{tt} u(\sigma,t) -V'(u(\sigma,t))},
\\ u(\sigma,t)= \pm 1, \quad|t|\ge \ell, \\ u(0,t) = x(t). 
\end{cases}
   $$
By standard theory, for each $x\in \mc X_\ell$ , we have that 
$S_\sigma(x)\restriction_\ell\in C^2((-\ell,\ell))$ for $\sigma>0$
and $\|S_\sigma(x)-x\|_\infty \to 0$ as $\sigma\to 0$. This implies
in particular 
that $\mc F_\ell(S_\sigma(x)) \le \mc F_\ell(x)$ for any $\sigma\ge 0$.
 By the
compactness of the level sets of $\mc F_\ell(x)$ there exists at least one
minimizer, say $\tilde x$. For what stated above we then have $\mc
F_\ell(S_\sigma(\tilde x)) = \mc F_\ell(\tilde x)$ for any $\sigma\ge 0$.
 By taking
the derivative we conclude that, for $\sigma>0$, $S_\sigma(\tilde x)$
 is a twice
differentiable solution of \eqref{ml}, hence $S_\sigma(\tilde x) =
m^*_\ell$. By continuity $\tilde x = m^*_\ell$.
\qed

\medskip
\noindent\emph{Proof of Theorem~\ref{gellconv}.}
The argument is rather standard and it is detailed below for completeness.
The mildly-equicoerciveness of $\mc F_\ell$ follows immediately from
Proposition~\ref{t:41}. We next prove the $\Gamma$-limsup inequality.
Given $x \in \mc X$ we define 
   $$
x_\ell(t) := \begin{cases} - 1 & \text{ if } t \in (-\infty, -\ell), \\
t+\ell-1 + x(-\ell+1)(t+\ell) 
& \text{ if } t\in [-\ell,-\ell+1], \\
x(t) & \text{ if } t\in (-\ell+1,\ell-1), \\
t-\ell+1 + x(\ell-1) (\ell-t) & \text{ if } t\in [\ell-1,\ell], \\
1 & \text{ if } t \in (\ell,+\infty).
\end{cases}
   $$
Clearly, $x_\ell \to x$ in $\mc X$. Moreover it is straightforward to
check that $\mc F_\ell(x_\ell) \to \mc F(x)$ since
the contribution in the set $[-\ell,-\ell+1] \cup [\ell-1,\ell]$
vanishes as $\ell\to\infty$ because $x(\pm(\ell-1)) \to \pm 1$ as
$\ell\to\infty$. We finally prove the $\Gamma$-liminf inequality. 
Pick $x\in \mc X$ and a sequence $x_\ell \to x$; if $\mc F_\ell (x_\ell) 
< +\infty$ we have $x_\ell(t) = \sgn(t)$ for $|t|\ge \ell$. Hence
$\mc F_\ell(x_\ell) = \mc F(x_\ell)$ and we conclude by
the lower semicontinuity of $\mc F$, which is established by noticing
that $\mc F(x) = \sup_{\ell}\: \int_{-\ell}^\ell\!dt\,\big[
\frac 12 x'(t)^2 + 2V(x(t))\big]$. 

To prove the last statement we first show that $\mc F(x)\ge\mc
F(\bam)$ for any $x\in \mc X$. Indeed, using the inequality
$a^2+b^2 \ge 2|ab|$ we have
   \begin{eqnarray*}
\mc F(x) & \ge & 2\int_{-\infty}^{+\infty}\!\!dt\, |x'(t)|\, \sqrt{V(x(t))}  \, \ge  \, 2 \, 
\Big| \, \int_{-\infty}^{+\infty}\!\!dt\, x'(t)\, \sqrt{V(x(t))} 
\, \Big |
\\ & = & 2 \int_{-1}^{+1}\!\!dy\, \sqrt{V(y)} = \int_{-1}^{+1}\!\!dy\, (1-y^2) = \frac 43 = \mc F(\bam).
   \end{eqnarray*}
On the other hand, in the above computation we get an equality if and
only if $|x'(t)| = 2 \sqrt{V(x(t))} = |1-x^2(t)|$. Since $x(\pm\infty)
=\pm 1$, this implies $x= \bam_z$ for some $z \in \bb R$. 
\qed

\smallskip
\noindent\emph{Proof of Theorem~\ref{barfell}.}
It is convenient to introduce the notation
\begin{equation}
\label{Fab}
\mc F_{[a,b]}(x) \, := \, \int_a^b \!dt\, 
\Big[ \frac 12\, x'(t)^2 + 2V(x(t)) \Big].
\end{equation}
The equi-coercivity of $\mc G_\ell$ is proven in Lemma~\ref{t:A4}.
We next prove the $\Gamma$-limsup inequality. By \eqref{glim},
it is enough to consider $x\in\mc M$. Recall that $m^*_\ell$ 
is the minimizer of $\mc F_\ell$ and note that, by the symmetry of
$V$, $m^*_\ell(0)=0$. Given $z\in \bb R$, for $\ell
\ge |z|$ we define
   \begin{equation}
   \label{recse}
m_z^{(\ell)}(t) := \begin{cases} -1 & \text{ if } t \in (-\infty, -\ell],
\\ m^*_{\ell+z}(t-z) & \text{ if } t \in (-\ell,z], \\
m^*_{\ell-z}(t-z) & \text{ if } t\in (z, \ell], \\
1 & \text{ if } t \in (\ell +\infty). \end{cases}
   \end{equation}
From \eqref{stima*} we get $m_z^{(\ell)} \to \bam_z$ in $\mc X$. We
claim that $m_z^{(\ell)}$ is a recovery sequence, i.e.\
   \begin{equation}
   \label{ub}
\lim_{\ell\to \infty} \mc G_\ell\big(m_z^{(\ell)}\big) =
16 \big[ \varch(4z) -1\big].
   \end{equation}
Indeed we have
   \begin{eqnarray*}
\mc F_\ell\big(m_z^{(\ell)}\big) & = & 
\mc F_{[-\ell,z]}\big(m_z^{(\ell)}\big) +
\mc F_{[z,\ell]}\big(m_z^{(\ell)}\big) \; = \; 
\mc F_{[-\ell-z,0]}\big(m^*_{\ell+z}\big) +
\mc F_{[0,\ell-z]}\big(m^*_{\ell-z}\big) \\
& = & \frac 12 \Big[ \mc F_{\ell+z}\big(m^*_{\ell+z}\big) +
\mc F_{\ell-z}\big(m^*_{\ell-z}\big)\Big],
   \end{eqnarray*}
where in the first step we used the translation covariance of 
$\mc F_{[a,b]}$ while, in the second one, that $t\mapsto m^*_\ell(t)$
is an odd function and $x\mapsto V(x)$ is an even function. Therefore
\begin{equation*}
\mc G_\ell\big(m_z^{(\ell)}\big) =
\frac{e^{4\ell}}2  \Big[\mc F_{\ell+z}\big(m^*_{\ell+z}\big)
- \frac 43 \Big] + \frac{e^{4\ell}}2 
\Big[\mc F_{\ell-z}\big(m^*_{\ell-z}\big) - \frac 43 \Big]  
+ e^{4\ell} \Big[\frac 43 - \mc F_\ell \big(m^*_\ell\big) \Big],
 \end{equation*}
and \eqref{ub} follows from \eqref{stima**}. 

We finally prove the $\Gamma$-liminf inequality. Let $x\in \mc X
\setminus  \mc M$, from Theorem~\ref{gellconv} and \eqref{stima**} it 
follows that, for any sequence $x_\ell\to x$, we have 
   $$
\varliminf_{\ell\to \infty}\;\mc F_\ell (x_\ell) \ge
\mc F(x) > \frac 43 = \lim_{\ell\to\infty} \mc F_\ell(m^*_\ell)
   $$
whence $\mc G_\ell(x_\ell) \to +\infty$ as $\ell\to\infty$. It remains 
to show that for any $z\in\bb R$ and any sequence $x_\ell\to\bam_z$ 
we have
   \begin{equation}
   \label{lig}
\varliminf_{\ell\to\infty} \, \mc G_\ell(x_\ell) \ge 
16 \big[ \varch(4z) -1\big].
   \end{equation}
It suffices to consider sequences
$x_\ell\to\bam_z$ such that $x_\ell(t) = \sgn(t)$ for $|t|
\ge \ell$. We next remark that, by
symmetry, the function $m^*_\ell(t)$, $t\in [0,\ell]$, is the unique
minimizer of $\mc F_{[0,\ell]}(x)$ with the boundary conditions
$x(0)=0$, $x(\ell)=1$. Analogously, $m^*_\ell(t)$, $t\in [-\ell,0]$,
is the unique minimizer of $\mc F_{[-\ell,0]}(x)$ with the boundary
conditions $x(-\ell)=-1$, $x(0)=0$. Since $x_\ell\to\bam_z$ we can
find a sequence $z_\ell \to z$ such that $x_\ell(z_\ell) =
0$. Recalling \eqref{recse}, by the translation covariance of
$\mc F_{[a,b]}$ we get
   \begin{eqnarray*}
\mc F_\ell(x_\ell) & = & 
\mc F_{[-\ell,z_\ell]}(x_\ell) + \mc F_{[z_\ell,\ell]}(x_\ell) 
\\ & \ge & \mc F_{[-\ell-z_\ell,0]}\big(m^*_{\ell+z_\ell}\big) + 
\mc F_{[0,\ell-z_\ell]}\big(m^*_{\ell-z_\ell}\big)
\; = \; \mc F_\ell(m_{z_\ell}^{(\ell)}).
   \end{eqnarray*}
The proof of \eqref{lig} is completed by observing
 that \eqref{ub} holds
also if the sequence $m_z^{(\ell)}$ is replaced by
$m_{z_\ell}^{(\ell)}$ with $z_\ell\to z$. 
\qed

   \section{Euclidean semiclassical approximation} 
   \label{sec:5}

From now on we set $\ell = \frac 14 \log\epsilon^{-1}$ and drop the 
subscript $\ell$ from the notation. We suppose given a filtered 
probability space $(\Omega, \mc S, \mc S_t, P)$ 
equipped with a standard Brownian motion $w_t$, $t\in [-\ell,\ell]$, 
with $w_{-\ell}=0$. By e.g.\ \cite[\S~5.6.B]{KS}, the Brownian bridge 
with diffusion coefficient $\epsilon$, starting at time $-\ell$ 
from $-1$ and arriving at time $\ell$ to $+1$, can be realized as 
the solution to the stochastic differential equation
   \begin{equation}
   \label{5p0}
\left\{\begin{array}{l}
d\eta_t = 
{\displaystyle \frac{1-\eta_t}{\ell-t}\, dt + \sqrt\epsilon\, dw_t}, \\
\eta_{-\ell} =  - 1, \end{array} \right. 
   \end{equation}
for $t\in [-\ell,\ell)$, and $\eta_t = \sgn(t)$ for $|t|\ge \ell$. 
Note in fact that the solution to the above equation satisfies 
$\lim_{t\uparrow\ell}\eta_t=1$ almost surely. 

Recalling the definition \eqref{S=}, given $\ell>0$ and  
$(t,x)\in [-\ell,\ell)\times \bb R$ we set 
   \begin{equation}
   \label{b=}
b(t,x) := - \partial_x S(\ell-t, x) = \begin{cases}
+ \sqrt{4V(x)+E_{t,x}} & \text{ if } x < 1 \\
0 & \text{ if } x = 1 \\ 
-\sqrt{4V(x)+E_{t,x}} & \text{ if } x > 1
\end{cases}
   \end{equation}
where $E_{t,x}$ is such that  
   \begin{equation}
   \label{Etx=}
\ell-t = \bigg| \int_x^1\! \frac{du}{\sqrt{4V(u) + E_{t,x}}}\bigg| 
   \end{equation}
The last equality in \eqref{b=} can be seen to follow for instance 
from \eqref{Etx=} and \eqref{forS}. 

We then define the process $\xi$ as the solution to the one 
dimensional stochastic differential equation
   \begin{equation}
   \label{5p1}
\left\{\begin{array}{l}
d\xi_t = b(t,\xi_t)\, dt + \sqrt\epsilon\, dw_t, \\
\xi_{-\ell} = - 1, \end{array} \right.
   \end{equation}
for $t\in [-\ell,\ell)$ and $\xi_t=\sgn(t)$ for $|t|\ge \ell$. 
We shall denote by $\nu_\epsilon$ the law of $\xi$.
Note that $b(t,x)>0$ for $x<1$ while $b(t,x)<0$ for $x>1$; moreover
$b(t,x)$ diverges as $t\uparrow\ell$ (unless $x=1$). Therefore
the drift in \eqref{5p1} drives the process $\xi$ from $-1$ at
time $-\ell$ to $1$ at time $\ell$. Finally, for $\ell$ large
and $(t,x)$ in compacts, $E_{t,x}\to 0$, so that we expect the solution
to \eqref{5p1} to converge, in the diagonal limit $\epsilon\to 0$
and $\ell\to +\infty$, to some $\bam_z\in\mc M$ which solves
$\dot x = \sqrt{4V(x)}$. We emphasize that in this limit some 
randomness will remain, as small deviation of the random force
affects the choice of $z$. Note indeed that $b(-\ell,-1) = 
\sqrt{E_{-\ell,-1}} = O(e^{-2\ell})$, which is of the same order
of the noise.
The above picture will be substantiated in the following. The required
analysis is not completely standard as it involves the joint limit 
$\epsilon\to 0$ and $\ell\to \infty$, and depends crucially on the
precise scaling $\ell = \frac 14 \log\epsilon^{-1}$.  

Before analyzing the process $\xi$ in itself, we show how it can be
used in the study of the $\phi^4_1$-measure with Dobrushin boundary
conditions. Recalling \eqref{S=}, we let 
   	 \begin{equation}
   \label{S0=}
S_0(t,x) := S(t,x) - \frac{(1-x)^2}{2t}.
   	\end{equation}
Note that $\frac{1}{2t}(1-x)^2$ is the action of a free particle, 
i.e.\ the infimum in \eqref{S=} in the case $V=0$. Since $S(t,x)$ 
satisfies the Hamilton-Jacobi equation with Hamiltonian $H(x,p) = 
\frac 12 p^2 - 2V(x)$, we get $S_0$ satisfies the equation
  \begin{equation}
   \label{HJS0}
\partial_t S_0(t,x) + \frac 12 \big[\partial_x S_0(t,x)\big]^2 - \frac{1-x}t \partial_x S_0(t,x) -2 V(x) = 0.
   \end{equation}
We also define $b_0(t,x) := -\partial_x S_0(\ell-t,x)$, so that 
$b(t,x) = b_0(t,x) + \frac{1-x}{\ell-t}$. It is shown in 
Theorem~\ref{t:31} that $S_0(t,x)$ is regular as $t\downarrow 0$,
therefore the drifts in \eqref{5p0} and \eqref{5p1} have the same
singular part. Indeed, we next show that $\xi$ is absolutely continuous 
w.r.t.\ the Brownian bridge $\eta$ and we obtain an explicit expression for the density of $\mu_\epsilon$ w.r.t.\ $\nu_\epsilon$. Recall the
definition \eqref{fell} of $\mc F_\ell$, and its minimizer  
$m^*_\ell$ considered in Proposition \ref{t:41}.

   \begin{proposition}
   \label{t:61}
We have
   \begin{equation}
   \label{6p1}
\frac{d\mu_\epsilon}{d\nu_\epsilon} (x) =
A_\epsilon^{-1} 
\: \exp\Big\{ - \frac 12 \int_{-\ell}^\ell\!dt\, 
\partial_{xx} S_0(\ell-t,x(t)) \Big\},
   \end{equation}
where 
   $$
A_\epsilon := \nu_\epsilon \Big(
e^{ - \frac 12 \int_{-\ell}^\ell\!dt\, 
\partial_{xx} S_0(\ell-t,x(t))} \Big)
=
Z_\epsilon \, e^{\epsilon^{-1} [\mc F_\ell(m^*_\ell) - \ell^{-1}]}.
   $$
   \end{proposition}

In Section \ref{sec:67} we shall analyze the measure 
$\mu_\epsilon$ as a perturbation of $\nu_\epsilon$.
Notice indeed that, while 
$\frac{d\mu_\epsilon}{d\rho_\epsilon} = e^{O(\epsilon^{-1}\ell)}$,
we have $\frac{d\mu_\epsilon}{d\nu_\epsilon} = e^{O(\ell)}$.

\proof Let $\psi_\ell : \mc X_\ell \to \mc X_\ell$ be the map 
defined as follows. The function $y = \psi_\ell(x)\restriction_\ell$ 
is the unique solution to
   \begin{equation}
   \label{5p4}
y(t) = x(t) + \int_{-\ell}^t\!ds\, \frac{\ell-t}{\ell-s}\, b_0(s,y(s)),
\qquad t\in [-\ell,\ell).
   \end{equation}
By writing the integral form of \eqref{5p1} and using Duhamel formula w.r.t.\ \eqref{5p0} we get that $\nu_\epsilon = \varrho_\epsilon 
\circ \psi_\ell^{-1}$. In particular the process $\xi$ is well defined
and satisfies $\lim_{t\uparrow \ell}\xi_t = 1$ almost surely. 
This representation of $\nu_\epsilon$, together with the regularity
of $b_0(t,x)$ proven in Theorem~\ref{t:31}, allows, by a standard truncation of which we omit the details, to use Girsanov theorem to obtain an explicit expression for the Radon-Nykodim derivative 
$\frac{d \nu_\epsilon}{d \varrho_\epsilon}$. We get
    \begin{equation}
    \label{5p2}
\frac{d \nu_\epsilon} {d \varrho_\epsilon}(\eta)  =
\exp\Big\{ \frac{1}{\sqrt\epsilon} \int_{-\ell}^\ell\! b_0(t,\eta_t)\, dw_t
- \frac{1}{2\epsilon} \int_{-\ell}^\ell\! dt \, b_0(t,\eta_t)^2 \Big\}.
   \end{equation}
On the other hand, by Ito's formula,
   \begin{eqnarray}
   \label{5p3}
&& \!\!\!\!\!\!\!\!\! S_0(0,\eta_\ell) - S_0(2\ell,\eta_{-\ell}) 
\nn \\ && \!\!\! 
= \int_{-\ell}^\ell\! dt \, \Big[ - \partial_t S_0(\ell - t,\eta_t)
+ \partial_x S_0 (\ell-t,\eta_t) \, \frac{1-\eta_t}{\ell-t} 
+ \frac \epsilon 2 \partial_{xx} S_0 (\ell-t,\eta_t) \Big]
\nn  \\ && \!\!\! + \, 
\sqrt\epsilon \int_{-\ell}^\ell\! \partial_x S_0(\ell-t,\eta_t) \, dw_t
   \end{eqnarray}
We note that $S(2\ell,-1) = \mc F_\ell(m^*_\ell)$, whence 
$S_0(2\ell,-1) = \mc F_\ell(m^*_\ell) -\frac 1\ell$ and $S_0(0,1)=0$.
Recalling $b_0(t,x) = - \partial_x S_0(\ell-t,x)$, by plugging 
\eqref{5p3} into \eqref{5p2} we get
   \begin{eqnarray*}
&&  \!\!\!\!\!\!\!\!\!\!\!\!
\frac{d \nu_\epsilon} {d \varrho_\epsilon} (\eta)
 \, = \, \exp\Big\{\epsilon^{-1}\big[ \mc F_\ell(m^*_\ell) - \ell^{-1}
 \big] + \frac 12 \int_{-\ell}^\ell\! dt \, 
\partial_{xx} S_0 (\ell-t,\eta_t)
\\ && - \epsilon^{-1} \int_{-\ell}^\ell\! dt \,
\Big[\partial_t S_0(\ell - t,\eta_t) - 
\partial_x S_0 (\ell-t,\eta_t) \, \frac{1-\eta_t}{\ell-t} 
+ \frac 12 \big[\partial_x S_0(\ell-t,\eta_t)\big]^2 \Big] \Big\}
\\ && \, = \, \exp\Big\{\epsilon^{-1} \big[\mc F_\ell(m^*_\ell) - \ell^{-1} 
\big]+ \frac 12 \int_{-\ell}^\ell\! dt \, \partial_{xx} S_0 (\ell-t,\eta_t) - 
\epsilon^{-1} \int_{-\ell}^\ell\! dt \, 2V(\eta_t) \Big\},
   \end{eqnarray*}
where in the last equality we used \eqref{HJS0}. Recalling \eqref{p2}, 
the identity \eqref{6p1} is thus proven. 
\qed

\medskip
We now turn to the analysis of \eqref{5p1}. Given $x\in \mc X_\ell$, 
let $\mc Z(x)$ be the leftmost zero of $x$, that is
\begin{equation}
\label{mcZ:} 
\mc Z(x):=\inf\{t\in [-\ell,\ell] \, : \: x(t) = 0\}.
\end{equation}
In the next theorem, whose proof is the main content of the present
section, we estimate the probability that the process $\xi$ lies in 
a small neighborhood of $\mc M$ and $\mc Z(\xi)$ stays in a compact.  

   \begin{theorem}
   \label{t:ap}
There exists $\eta_0>0$ such that, for any $\eta\in (0,\eta_0)$ the following holds. There exist positive reals $a_0$ and  $\epsilon_0$ 
such that, for any $\epsilon\in (0,\epsilon_0]$ and $a 
\in[a_0,\ell]$ we have
   \begin{eqnarray}
   \label{5.8}
&& \nu_\epsilon\big(\big\{x\, :\: d(x,\mc M) > \epsilon^{\frac
12 -\eta} \big\}\big) \le \exp\big\{-\epsilon^{-\frac 12 \eta}\big\},
\\ &&  \label{pa1}
\nu_\epsilon\big(\big\{x\, :\: |\mc Z(x)| > a 
\big\}\big) \le \exp\big\{-\epsilon^{-\frac 12 \eta}\big\}
+ 2\exp\big\{-e^{\frac a4}\big\},
\\ && \label{e:P}
\nu_\epsilon\big(\big\{\exists\, t\in [-\ell+a,a] \, : \:
x(t) < \bam_a(t) \big\} \big) \le \exp\big\{-e^{\frac a4} \big\}.
   \end{eqnarray}
   \end{theorem}
   
By the aforementioned behavior of $b(t,x)$, since the noise is of
order $\sqrt\epsilon$ and the time interval is of order 
$\log\epsilon^{-1}$, the process $\xi$ will essentially move inside
the interval $[-1,1]$. The precise statement is the following. 

   \begin{lemma}
   \label{t:52}
For each $\delta>0$ we have 
   $$
\nu_\epsilon\Big(\sup_{t\in\bb R} |x(t)| > 1 + \delta \Big)
\le \frac8{\sqrt{\pi}} \frac{\sqrt{\epsilon\ell}}{\delta} \,
\exp\Big\{-\frac{\delta^2}{16\epsilon\ell}\Big\}.
    $$
   \end{lemma}

\proof 
Let $\xi$ be the solution of \eqref{5p1}, and  introduce the event 
   $$
\mc B := \Big\{\sup_{t\in [-\ell,\ell]} |w_t| < 
\frac{\delta}{2\sqrt\epsilon}\Big\}.
   $$ 
By the reflection principle, 
\begin{equation}
\label{stB}
P(\mc B^\complement) \le 4 P\Big(w_{\ell} \ge
\frac{\delta}{2\sqrt\epsilon} \Big) \le \frac8{\sqrt{\pi}}
\frac{\sqrt{\epsilon\ell}}{\delta} \,
\exp\Big\{-\frac{\delta^2}{16\epsilon\ell}\Big\}.
\end{equation}

We claim that $\sup_{t\in [-\ell,\ell]} |\xi_t| < 1 + \delta$ on 
the event $\mc B$. We shall only prove that $\inf_{t\in [-\ell,\ell]} 
\xi_t > - 1 - \delta$, a symmetric argument shows that we also have
$\sup_{t\in [-\ell,\ell]} \xi_t < 1 + \delta$.
Let $\tau$ be the first time $\xi_t$ hits $-1-\delta$. If there is no
such $\tau$ in the interval $(-\ell,\ell)$ we are done. Otherwise let
$\sigma<\tau$ be the last passage by $-1$ before $\tau$. By
integrating \eqref{5p1} in the time interval $[\sigma,\tau]$ and using
that $b(t,x) \ge 0$ for $(t,x) \in (-\ell,\ell) \times (-\infty,-1]$,
see \eqref{b=}, we get
   $$
- \delta = \xi_\tau-\xi_\sigma =  \int_\sigma^\tau\!ds\, b(s,\xi_s) 
+ \sqrt\epsilon \big(w_\tau-w_\sigma\big) \ge - 2 \sqrt\epsilon\, 
\sup_{t\in [-\ell,\ell]} |w_t|,
   $$
which gives a contradiction.
\qed

\smallskip
In the following lemma we show that, for $t$ away from the boundary,
the solution of \eqref{5p1} is in a small neighborhood of some
profile $\bam_z\in\mc M$ with probability close to one. We also 
identify $z$ as a zero (it does not matter which one) of 
$t\mapsto\xi_t$.
  
   \begin{lemma}
   \label{t:53}
For each $\eta\in \big(0, \frac 12 \big)$ and $\sigma\in
(0,1-2\eta)$ there exists $\epsilon_0$ such that for any
$\epsilon \in (0, \epsilon_0]$ we have
   \begin{equation}
   \label{tube}
\nu_\epsilon\Big(|\mc Z(x)| < \sqrt\ell\, ; \,
\sup_{|t-\mc Z(x)| \le \sigma \ell} \big| x(t) -
\bam_{\mc Z(x)}(t)\big| > \epsilon^{\frac 12 -\eta} \Big) \le 
\exp\big\{-\epsilon^{-\eta}\big\}.
   \end{equation}
   \end{lemma}

\proof We shorthand $\mc Z(x)$ by $z$ and define
\begin{eqnarray*}
\tau_- & := & \sup\big\{ t \le z \, : \: 
\big| x(t) - \bam_z(t)\big| > \epsilon^{\frac 12 -\eta} \big\}
\lor
(z-\sigma \ell), 
\\ \tau_+ & := & \inf\big\{ t \ge z  \, : \: 
\big|x(t) - \bam_z(t) \big| > \epsilon^{\frac 12 -\eta} \big\}\land 
(z + \sigma\ell).
\end{eqnarray*}
Let also
\begin{equation}
\label{B1}
\mc B_1 := \Big\{w\, : \: \sup_{t\in [-\ell,\ell]}|w_t| < 
\epsilon^{-\frac 23 \eta}\Big\}.
\end{equation}
By the bound \eqref{stB} we have $P\big(\mc B_{1}^\complement\big) \le 
\exp\big\{- \epsilon^{-\eta}\big\}$ for any $\epsilon$ small enough.
The proof will be completed  by showing that on the event 
$\mc B_{1}\cap \{|z|< \sqrt\ell\}$ we have $\tau_\pm = z\pm\sigma\ell$
for any $\epsilon$ small enough. 

Let $v_t := \xi_t - \bam_z(t)$. Integrating \eqref{5p1} and using 
$\bam'_{z}=1-\bam^2_z$
  we obtain that for $t\in [\tau_-,\tau_+]$, 
\begin{equation}
\label{eqv}
v_t = \int_z^t\!ds\, \Big[-2\bam_z(s) v_s + R(s,v_s) \Big] +
\sqrt\epsilon\, [w_t - w_z],
\end{equation}
with 
\begin{align}
\label{Rv}
R(s,y) & :=  b(s,\bam_z(s)+y) - [1- \bam_z^2(s)] +  2\bam_z(s)\,y
 \nn \\ 
& =  \sqrt{\big[1-(\bam_z(s)+y)^2\big]^2 + 
E_{s,\bam_z(s)+y}} - [1-(\bam_z(s)+y)^2] - y^2
\end{align}
where we used \eqref{b=},  
 with $\bam_z(s)+ v_s \in (-1,1)$, 
which holds for any $\epsilon$ small enough since,
 for $\eta\in \big[0,\frac 12\big)$, $\sigma\in (0,1-2\eta)$, and 
$s\in  [\tau_-,\tau_+]$,
$$
\big|\bam_z(s) + v_s\big| \le 1-e^{-2|s-z|} + \epsilon^{\frac 12 -\eta}
\le 1 - e^{-2\sigma\ell} + e^{-2\ell+4\eta\ell} < 1.
$$  
Integration of \eqref{eqv}, using that 
$-2 \bam_z = (\log\bam'_z)'$ and  $\bam'(t)
= \varch(t)^{-2}$ yields that,  on the event $\mc B_{1}$, 
\begin{eqnarray}
\label{v<=}
|v_t| & = & \bigg| \int_z^t\!ds\, \frac{\varch^2(s-z)}{\varch^2(t-z)}
\Big[R(s,v_s) - 2 \sqrt\epsilon \, \bam_z(s) \big(w_s-w_z\big)
\Big] + \sqrt\epsilon \big(w_t-w_z\big) \bigg| 
\nn\\ & \le & \bigg| \int_z^t\!ds\, \frac{\varch^2(s-z)}{\varch^2(t-z)}
R(s,v_s)\bigg| + 6\,  \epsilon^{\frac 12 -\frac 23 \eta}.  
\end{eqnarray}
where we used $\Big|\int_z^t\!ds\, \frac{\varch^2(s-z)}{\varch^2(t-z)}
\Big| \le 1$. 

We claim that, for each $\eta\in \big[0, \frac 12 \big)$ and 
$\sigma\in (0,1-2\eta)$ there exists a constant $C>0$ such that, 
for any $\epsilon$ small enough,
\begin{equation}
\label{EE}
\sup_{t\in [\tau_-,\tau_+]} E_{t,\bam_z(t) + v_t} \le C \ell^4 
e^{-4(\ell-z)}, 
\end{equation}
whose proof is given in Appendix~\ref{s:ap}.
By using \eqref{EE} in \eqref{Rv} we get, for $s\in  
[\tau_-,\tau_+]$, 
\begin{eqnarray*}
|R(s,v_s)| & \le & v_s^2 + \frac{E_{s,\bam_z(s) + v_s}}
{2\,\big[1- \big(\bam_z(s)+v_s\big)^2\big]}\le v_s^2 +
\frac{ C \ell ^4 e^{-4(\ell-z)}}{1 - \big(|\bam_z(s)| + 
\epsilon^{\frac 12 -\eta}\big)^2} 
\\ & \le &  v_s^2 + 2C\ell^4\, \exp\{-4(\ell-z) + 2 |s-z|\}, 
\end{eqnarray*}
where in the last inequality we used $1-|\bam_z(s)| \ge  e^{-2|s-z|}$
and $e^{-2|s-z|} > \epsilon^{\frac 12 -\eta}$ for $|s-z|\le
\sigma\ell$. Plugging this bound into \eqref{v<=} and using the 
estimate
$$
\frac{\varch^2(s-z)}{\varch^2(t-z)}\le 4\,e^{-2|t-s|} \qquad 
\text{for }  s\in [t\land z\,,\,t\lor z],
$$ 
we get that, on the event $\mc B_{1}\cap\{|z|<\sqrt\ell\}$, for any 
$t\in [\tau_-,\tau_+]$ and $\epsilon$ small enough,
\begin{eqnarray}
\label{v1}
|v_t| & \le & \bigg| \int_z^t\!ds\, 4\,e^{-2|t-s|}\, v_s^2 \bigg|
+ 2 C \ell^4 \, \exp\{-4\ell + 4z + 2 |t-z|\} + 6\,  
\epsilon^{\frac 12 -\frac 23 \eta}
\nn \\ & \le & \bigg| \int_z^t\!ds\, 4\,e^{-2|t-s|}\, v_s^2 \bigg|
+  2 C \ell^4 \, \sqrt\epsilon + 6 \, \epsilon^{\frac 12 -\frac 23 \eta}
\nn \\ & \le & \bigg| \int_z^t\!ds\, 4\,e^{-2|t-s|}\, v_s^2 \bigg|
+ 7 \, \epsilon^{\frac 12 -\frac 23 \eta},
\end{eqnarray}
where in the second inequality we used that $-4\ell + 4z + 2 |t-z| <
- 4 \ell + 4\sqrt\ell + 2\sigma\ell < -2\ell$ for $\eta\in \big[0,
\frac 12 \big)$, $\sigma\in (0,1-2\eta)$, and $\ell$ large enough. 
By \eqref{v1} and a standard bootstrap argument it follows that 
$\tau_\pm = z\pm\sigma\ell$ for any $\epsilon$ small enough. 
\qed

\smallskip
In order to complete the proof of Theorem~\ref{t:ap} we need
to analyze the behavior of $\xi_t$ for $t$ close to the boundaries.
We remark that while both the measures $\varrho_\epsilon$ and 
$\mu_\epsilon$ are invariant w.r.t.\ the map $x(t) \mapsto
-x(-t)$, this symmetry property does not hold for $\nu_\epsilon$.
We need therefore two separate arguments. We start with $t<0$
and, in the next  lemma, we give an upper bound for the probability 
that $\xi(t)$ gets above $\bam_{-a}(t)$ for $t\le-a$ and $a$ large. 

  \begin{lemma}
   \label{t:525}
There exist reals $\epsilon_0$ and $a_0>0$ such that, for any 
$a\in [a_0,\ell]$ and $\epsilon\in (0,\epsilon_0]$, we have
   $$
\nu_\epsilon\Big(\big\{x\, : \: \exists\, t\in [-\ell, -a] \text{ such that } x(t) > \bam_{-a}(t) \big\} \Big) \le  \exp\big\{- e^{\frac a4}
\big\}.
   $$
   \end{lemma}

\proof We introduce the event
   \begin{equation}
   \label{5pp-1}
\mc B_{1,a} := \Big\{w\, : \: |w_t| < e^{\frac a4} + (t + \ell) 
\;\;\;\; \forall\,t\in [-\ell,\ell]\Big\}.
   \end{equation}
The probability of $\mc B_{1,a}$ can be computed explicitly,
see e.g.\ \cite[\S~4.3.C]{KS}. We give however a short proof of
the bound  
   \begin{equation}
   \label{5pp0}
P(\mc B_{1,a}^\complement) \le 2\exp\big\{-2\, e^{\frac a4}\big\}.
   \end{equation}
Indeed, let $M_t := \exp\{2w_t-2(t+\ell)\}$, $t\in [-\ell,\ell]$.
Since $M_t$ is a mean one continuous martingale, by Doob inequality 
we have
\begin{eqnarray*}
&& P\Big(\big\{\exists\, t\in [-\ell,\ell]\, : \: w_t \ge e^{\frac a4} +
t + \ell \big\}\Big) 
\\ && \phantom{mmmmmmm} = \, P\Big(\sup_{t\in [-\ell,\ell]} M_t 
\ge \exp\big\{2\, e^{\frac a4} \big\}\Big)\le \exp\big\{-2\, 
e^{\frac a4} \big\}
\end{eqnarray*}
and the bound \eqref{5pp0} follows. 

We next show that there exist $\epsilon_0, a_0 >0 $ such that, for any 
$a\in [a_0,\ell]$ and $\epsilon\in (0,\epsilon_0]$, on the event 
$\mc B_{1,a}$ we have $\xi_t < \bam_{-a}(t)$ for any $t\in [-\ell, -a]$.
Let $\tau:=\inf\{t\ge-\ell: \xi_t = \bam_{-a}(t)\}\wedge (- a)$. 
Note that, by continuity $\tau>-\ell$; we show that $\tau=-a$ on
the event $\mc B_{1,a}$ arguing by contradiction. Indeed, let
$\sigma \in [-\ell,\tau)$ be the last time for which $\xi_t =
m^*_\ell(t)$, the minimizer defined in Proposition~\ref{t:41}.
We integrate the equation \eqref{5p1} in the interval $[\sigma,t]$ 
with $t\in [\sigma,\tau]$, getting 
   \begin{equation*}
\xi_t = \xi_\sigma + \int_\sigma^t\!ds\, b(s,\xi_s) + \sqrt\epsilon\,
(w_t-w_\sigma).
   \end{equation*}
Since for $s\in [\sigma,\tau]$ we have $-1\le m^*_\ell(s) \le \xi_s \le
\bam_{-a}(s) \le 0$, from \eqref{b=}, the inequality 
$\sqrt{\alpha+\beta} \le \sqrt \alpha +\sqrt \beta$, 
$\alpha,\beta\ge 0$, and \eqref{bm1} we have
   \begin{equation}
   \label{5pp1}
\xi_t \le m^*_\ell(\sigma) + \int_\sigma^t\!ds\, \Big[ 1-\xi_s^2 + 
\sqrt{E_{s,m^*_\ell(s)}}\,\Big] + \sqrt\epsilon\, (w_t-w_\sigma).
   \end{equation}      
Set $v_t := \xi_t - \bam_{-a}(t)$ and note that 
$E_{s,m^*_\ell(s)} = E_{-\ell,-1} = E_\ell$. 
Since $\bam_{-a}'(s) = 1 - \bam_{-a}^2(s)$ and 
   $$
1-\xi_s^2 - [1- \bam_{-a}(s)^2] = 
- 2 \bam_{-a}(s)v_s - v_s^2 \le - 2 \bam_{-a}(s)v_s,
   $$   
from \eqref{5pp1} and \eqref{El} we get that, 
for $\epsilon$ small enough, 
   \begin{equation}
   \label{5pp2}
v_t \le - \big[\bam_{-a}(\sigma) - m^*_\ell(\sigma)\big]
+ \int_\sigma^t\!ds\, \Big[ - 2 \bam_{-a}(s)v_s +
9 \sqrt\epsilon \Big] + \sqrt\epsilon\, (w_t-w_\sigma).
   \end{equation}
Next, we show  that, on the
event $\mc B_{1,a}$, we have $v_\tau <0$ provided $a$ is large enough
and $\epsilon$ is small enough, what contradicts the assumption 
$\tau\in [-\ell, -a)$. To this end we integrate the inequality 
\eqref{5pp2}, proceeding as explained when getting \eqref{v<=} from 
\eqref{eqv}, obtaining 
   \begin{eqnarray}
   \label{v<}
v_\tau & \le & - \,
\frac{\varch^2(\sigma+a)}{\varch^2(\tau+a)}
\big[ \bam_{-a}(\sigma) - m^*_\ell(\sigma)\big] + 
\sqrt\epsilon (w_\tau-w_\sigma) 
\nn \\ && 
+\, \sqrt\epsilon\,\int_\sigma^\tau\!dt\, 
\frac{\varch^2(t+a)}{\varch^2(\tau+a)}
\Big(9 - 2 \bam_{-a}(t) [w_t-w_\sigma] \Big).
   \end{eqnarray}
We now observe that, by \eqref{bm0i}, 
   $$
\bam_{-a}(\sigma) - m^*_\ell(\sigma) \ge \bam_{-a}(\sigma) - \bam_0 (\sigma) 
- A_1 \sqrt\epsilon \ge e^{2\sigma} \, \big(e^{2a}-2\big) - A_1 \sqrt\epsilon, 
   $$
where we used $e^{2a} \le 1+\varth(a) \le 2 e^{2a}$, $a\le 0$. On the
other hand, since  
\begin{equation}
  \label{eq:stker}
\frac 12\, e^{\beta-\alpha} \le \frac{\varch(\alpha)}{\varch(\beta)}
\le e^{\beta-\alpha}, \qquad  \alpha <\beta\le 0,  
\end{equation}
the inequality \eqref{v<} yields
   \begin{eqnarray*}
v_\tau & < & - \frac 14\, e^{2(\tau-\sigma)} \Big[ e^{2\sigma}
\big(e^{2a}-2\big) - A_1\sqrt\epsilon\Big] + 2 \sqrt\epsilon\, \big(
\tau+\ell+e^{\frac a4}\big) \\ && +\, \sqrt\epsilon \int_\sigma^\tau\!dt\,e^{2(\tau-t)}
\Big[9+4\big(t+\ell+e^{\frac a4}\big)\Big] 
\\ & \le & - \frac 14 \, \sqrt\epsilon\, e^{2(\tau+\ell)}\bigg\{
e^{2a} - 2 - A_1 - 8\, e^{-2(\tau+\ell)}(\tau+\ell + e^{\frac a4})
\\ && \phantom{ - \frac 14 \, \sqrt\epsilon\, e^{2(\tau+\ell)}\bigg\{}
- 4 \int_0^\infty\!ds\,e^{-2s} \Big[9+4\, (s+e^{\frac a4})\Big]\bigg\}.
   \end{eqnarray*}
By choosing $a_0$ large enough the term inside the curly brackets above
is strictly positive for any $a\ge a_0$. This yields $v_\tau<0$ which
is the  contradiction announced and, together with \eqref{5pp0}, 
concludes the proof of the lemma.  
\qed

\smallskip
The analysis for $t>0$ is somewhat more delicate. As a first step,
which is the content of the next lemma, we study the process $\xi_t$
for $t\in [-\ell+a,a]$ and show that for $a$ large it does not get below
$\bam_a(\cdot)$. In particular $\xi_a \ge 0$ with probability close
to one. In Lemma~\ref{t:526} below we then show that this property
yields an upper bound on the probability that $\xi_t$ gets below
$\bam_a(\cdot)$ for some $t\in [a,\ell]$. 

\begin{lemma}
\label{t:P}
There exist $a_0$, 
$\epsilon_0>0$ such that, for any $a\in \big[a_0,\ell\big]$ and $\epsilon\in (0,\epsilon_0]$, we have 
   $$
\nu_\epsilon\Big(\big\{\exists\, t\in [-\ell+a,a] \, : \:
x(t) < \bam_a(t) \big\} \Big) \le \exp\big\{-e^{\frac a4} \big\}.
   $$
   \end{lemma}

\proof The proof will be completed in three steps, each one taking place
with probability close to one for $a$ large and $\epsilon$ small. 
We first show that in the time interval $[-\ell,-\ell+a]$ the process 
$\xi_t$ reaches the level $-1+\sqrt{\epsilon}\, e^{\frac a3}$. 
We then show that $\xi_t$ hits the level 
$\epsilon^{\frac 38}$ before hitting $\bam_a(\cdot)$. Finally, once the
process is above $\epsilon^{\frac 38}$ does not go below zero. 

\noindent \emph{Step 1.} We introduce the event
   \begin{equation}
   \label{5pp-merda}
\mc B_{2,a} := \Big\{w\, : \: \sup_{t\in[-\ell,-\ell+a]} 
|w_t| < e^{\frac a3} \Big\}.
   \end{equation}
Note that, for $a$ large enough, we have $ P(\mc B_{2,a}^\complement) \le
\exp\big\{- e^{\frac a2} \big\}$. We claim that on the event $\mc B_{2,a}$
there exists a time $\tau_1\in [-\ell,-\ell+a]$ such that $\xi_{\tau_1}
\ge -1 + \sqrt\epsilon\, e^{\frac a3}$. We argue by contradiction.  If
there is no such $\tau_1$ we have $\xi_t < -1 + \sqrt\epsilon\, 
e^{\frac a3} \le 0$ for any $t\in [-\ell,-\ell+a]$
(we choose $\epsilon_0$ so small that $\sqrt \epsilon\,
e^{\frac\ell 3} \le 1$). From the inequality $(1+\xi_t)e^{\ell-t} \le
e^{\frac a3}$, \eqref{b=}, and \eqref{E>} we get 
$$
\xi_t \ge -1 + \sqrt\epsilon \int_{-\ell}^t\!ds\, 2 e^{s+\ell-\frac a3}
- \sqrt\epsilon e^{\frac a3} \ge -1 + 2 \sqrt\epsilon\, e^{-\frac a3} \big(
e^{t+\ell} - 1\big) - \sqrt\epsilon\, e^{\frac a3}.
$$
In particular, for $a$ large enough, $\xi_{-\ell+a} > -1 + 
\sqrt\epsilon\, e^{\frac a3}$ which contradicts the definition of 
$\tau_1$. 

\noindent \emph{Step 2.} 
Let $\tau_1 \in [-\ell,-\ell+a]$ be as in Step 1. We define 
$\tau_2:=\inf\{t>\tau_1 \,:\: \xi_t \le \bam_a(t)\}\land a$ and 
$\tau_3 := \inf\{t>\tau_1 \,:\: \xi_t \ge \epsilon^{\frac 38}\}\land a$.
On the event $\mc B_{2,a}$, for $a$ large enough we have
$\xi_{\tau_1} >\bam_a(\tau_1)$; hence $\tau_2, \tau_3 >\tau_1$. 
Consider  the event $\mc B_{1,a}$ that has been defined in 
\eqref{5pp-1}. We claim that, by taking $\epsilon$ small enough and 
$a$ large enough, on the event $\mc B_{2,a}\cap \mc B_{1,a}$ we have
$\tau_3<\tau_2$. We argue by contradiction, i.e.\ we assume that
$\xi_t<\epsilon^{\frac 38}$ for any $t\in (\tau_1,\tau_2]$. Let $T$ such that $\bam_a(T)=-\epsilon^{\frac 38}$ and set $v_t := \xi_t -\bam_a(t)$. 
Integrating \eqref{5p1} in the time interval $[\tau_1,t]$, with $t\in [\tau_1,\tau_2\land T]$, using \eqref{b=} and $\bam'_a=1-\bam_a^2$ 
we get
\begin{eqnarray*}
  v_t &\ge & \xi_{\tau_1} - \bam_a(\tau_1) + 
 \int_{\tau_1}^t\!ds\, \big[ -\bam_a(s) - \xi_s] \, v_s
  + \sqrt\epsilon\,[w_t-w_{\tau_1}] \\
&\ge & -1+\sqrt\epsilon\, e^{\frac a3} - \bam_a(-\ell+a) + 
  \int_{\tau_1}^t\!ds\, \Big[ -\bam_a(s) -\epsilon^{\frac 38}\ \Big] \, v_s
  + \sqrt\epsilon\,[w_t-w_{\tau_1}].
\end{eqnarray*}

Note that $\bam_a(-\ell+a)\le -1 + 2 \sqrt\epsilon $. By integrating
the above inequality, proceeding as in \eqref{v<}, we thus find
\begin{eqnarray*}
  v_t & \ge & 
  \frac{\varch(\tau_1-a)}{\varch(t-a)} e^{-\epsilon^{\frac 38} (t-\tau_1)} 
  \sqrt\epsilon\big(e^{\frac a3} - 2 \big) 
    + \sqrt\epsilon [ w_t-w_{\tau_1}] 
    \\
    &&+\;\sqrt\epsilon \int_{\tau_1}^t\!ds \,
    \frac{\varch(s-a)}{\varch(t-a)} e^{-\epsilon^{\frac 38} (t-s)}\Big[
    -  \bam_a(s) - \epsilon^{\frac 38}\Big] [w_s-w_{\tau_1}],
\end{eqnarray*}
whence, by \eqref{eq:stker} and the definition of the event $\mc
B_{1,a}$ (we suppose $\epsilon$ so small that $e^{-\epsilon^{\frac 38}}
(t-\tau_1) > \frac 12$),
\begin{eqnarray*}
  v_t & > & \frac 14\, \sqrt\epsilon\, e^{ (t-\tau_1)} 
  \bigg\{
  e^{\frac a3} - 2 - 8 e^{-(t-\tau_1)} 
  \big[e^{\frac a4}+(t+\ell)\big]   
  \\
  &&\phantom{ \frac 14 \sqrt\epsilon e^{-(t-\tau_1)} \bigg\{}
  - 9 \int_{\tau_1}^t\!ds\, e^{- (s-\tau_1)} \, 
  \big[e^{\frac a4}+s+\ell)\big]   \bigg\} \\
  &\ge & \frac 14\, \sqrt\epsilon \, e^{(t-\tau_1)} 
  \Big\{
  e^{\frac a3}- 2 - 17 \big[e^{\frac a4} + a + 1 \big]
  \Big\},
\end{eqnarray*}
where we used that $\tau_1+\ell\le a$.
By choosing $a_0$ large enough we get that the term inside
the curly brackets above is strictly positive for any $a\ge a_0$, so   
$\tau_2\land T=T$. Finally, by evaluating the above inequality for 
$t=T$, we conclude that $\xi_T>4\epsilon^{\frac 38}$ which gives the desired
contradiction. 

\noindent \emph{Step 3.}
Let $\tau_3$ be as in Step 2. We claim that on the event
$\mc B_{2,a}\cap \mc B_{1,a}$ we have $\xi_t \ge 0$ for any $t\in
[\tau_3,a]$. Assume this is not the case, let $\sigma_+>\tau_3$ be
the hitting time of the level zero, and let $\sigma_- :=
\sup\{t\in [\tau_3,\sigma_+) \, : \, \xi_t = \epsilon^{\frac 38}\}$. 
By using \eqref{b=} for $t\in [\sigma_-,\sigma_+]$ we have
\begin{equation*}
\xi_t \ge \epsilon^{\frac 38} + \sqrt\epsilon \big(w_t -w_{\sigma_-}\big).
\end{equation*}
Recalling \eqref{5pp-1} this gives the contradiction $\xi_{\sigma_+}>0$.
\qed

\begin{lemma}
\label{t:526}
For each $\eta \in \big(0,\frac 12)$ there exist  $a_0$, 
$\epsilon_0>0$ such that,  for any $a\in \big[a_0,\ell\big]$ and 
$\epsilon\in (0,\epsilon_0]$, we have 
   $$
\nu_\epsilon\Big(\big\{\exists\, t\in [a,\ell] \, : \:
x(t) < \bam_a(t)-\epsilon^{\frac 12-\eta}\big\} \Big)
\le \exp\big\{-\epsilon^{-\eta}\big\} + \exp\big\{-e^{\frac a4}
\big\}.
   $$
   \end{lemma}

\proof By Lemma~\ref{t:P}, for 
$ a\ge a_0$,  $\nu_\epsilon\big(\big\{x\, : \: 
x(a) < 0  \big\} \big) \le  \exp\big\{- e^{\frac a4}\big\}$. 
Let $\tau:=\inf\{t\in[a,\ell]\,:\: \xi_t\ge
1\}\land \ell $. Recalling $\mc B_1$ is defined in \eqref{B1},
 we claim that on the event $\mc B_1\cap \{\xi_a \ge 0\}$ we have
$\xi_t \ge \bam_a(t) -\epsilon^{\frac 12 -\eta}$ for any 
$t\in[a,\tau]$. If $\tau = a$ there is
nothing to prove, otherwise let $\varkappa:= \xi_a\in[0,1]$ and define
$\gamma_t$, $t\in [a,\tau]$, as the solution to 
\begin{equation*}
\gamma_t = \varkappa + \int_a^t\!ds (1-\gamma_s^2) + \sqrt\epsilon
\big(w_t - w_a).
\end{equation*}
By \eqref{b=} we have $\xi_t \ge \gamma_t$. Letting 
$a_\varkappa:= a-\bam_0^{-1}(\varkappa)$ we note 
$\bam_{a_\varkappa}(t)$, $t\in[a,\tau]$, solves 
\begin{equation*}
\bam_{a_\varkappa}(t) = \varkappa + \int_a^t\!ds\, 
\big[1-\bam_{a_\varkappa} (s)^2\big].
\end{equation*}
By setting $v_t = \gamma_t - \bam_{a_\varkappa}(t)$ and proceeding as in 
\eqref{v<=}, on the event $\mc B_1\cap \mc B_{2,a}\cap \mc B_{1,a}$,
see \eqref{5pp-1} and \eqref{5pp-merda}, we get 
\begin{equation*}
|v_t| \le \int_a^t\!ds\, 
\frac{\varch^2(s-a_\varkappa)}{\varch^2(t-a_\varkappa)}
\, v_s^2 + 10\, \epsilon^{\frac 12 -\frac 23\eta}.
\end{equation*}
By a  standard bootstrap argument we deduce that $\sup_{t\in[a,\tau]}
|v_t| \le \epsilon^{\frac 12 -\eta}$ for $\epsilon$ small enough. 
This concludes the proof of the claim. 

Finally, since $b(t,x) \ge 0$ for any $(t,x)
\in [-\ell,\ell] \times (-\infty,1]$, by the same argument given in
the proof of  
Lemma~\ref{t:52} we get that, for each $\eta\in\big(0, \frac 12\big)$
and $\epsilon$ small enough,
$$
P \Big(\big\{\exists\, t\in [\tau,\ell] \, : \: \xi_t < 1 -
\epsilon^{\frac 12 -\eta} \big\} \Big) \le 
\exp\big\{-\epsilon^{-\eta}\big\},
$$
which concludes the proof.
\qed

\medskip
We have now collected all the ingredients needed to conclude
the proof of the main result of this section. 

\smallskip
\noindent\emph{Proof of Theorem~\ref{t:ap}.} Recall that $\mc Z(x)$
denotes the leftmost zero of $x\in\mc X_\ell$. The bound \eqref{pa1}
follows directly from Lemmata~\ref{t:525} and \ref{t:526}.
The bound \eqref{e:P} is the content of Lemma~\ref{t:P}. In order
to prove \eqref{5.8}, by \eqref{pa1}, it is enough to prove the
following. For each $\eta$ small enough there exists $\epsilon_0>0$ 
such that, for any $\epsilon \in (0,\epsilon_0]$, we have
\begin{equation}
\label{pa2}
\nu_\epsilon\Big(\sup_{t\in[-\ell,\ell]} \big| x(t) -
\bam_{\mc Z(x)}(t) \big| > \epsilon^{\frac 12-\eta},\,\, |\mc Z(x)| \le 
a_\ell \Big) \le \frac 12 \exp\big\{\epsilon^{-\frac 12 \eta}\big\},
\end{equation}
where $a_\ell:= \log^2\ell$.

We consider separately the cases
$t\in [-\ell,-\sigma\ell]$, $t\in [\sigma\ell,\ell]$, and
$t\in [-\sigma\ell,\sigma\ell]$, with $\sigma <1 $ suitably chosen. 
For the first case, we observe that by Lemma~\ref{t:52} with 
$\delta=\frac 12 \epsilon^{\frac 12 -\eta}$ and Lemma~\ref{t:525}, 
for any $\epsilon$ small enough,
\begin{equation*}
\nu_\epsilon\Big( \exists\, t\in [-\ell,-\sigma\ell] \, : \:
x(t) \notin \big(-1-{\textstyle\frac 12} \epsilon^{\frac 12 -
  \eta},\bam_{-a_\ell}(t)\big) \Big) \le 
  \exp\big\{-\epsilon^{-\eta}\big\}
+ \exp\big\{-e^{\frac 14 a_\ell}\big\}.
\end{equation*}
We observe that $\bam_{-a_\ell}(t) \le -1 + 2 
\epsilon^{\frac\sigma 2} e^{2a_\ell}$
for any $t\in [-\ell,-\sigma\ell]$. By choosing 
$\sigma\in[\sigma_0,1)$ with $\sigma_0 := 1- \frac 53 \eta$,
 the previous estimate implies
\begin{eqnarray*}
&& \nu_\epsilon\Big(\sup_{t\in[-\ell,-\sigma\ell]} \big| x(t) -
\bam_{\mc Z(x)}(t) \big| > \epsilon^{\frac 12-\eta}, |\mc Z(x)| \le 
a_\ell \Big) \nn \\ && \phantom{mmmmmm} \le 
\exp\big\{-\epsilon^{-\eta}\big\} + \exp\big\{-e^{\frac 14 a_\ell}
\big\}.
\end{eqnarray*}
Analogously, for the second case, by Lemma~\ref{t:52} with
$\delta=\frac 12 \epsilon^{\frac 12 -\eta}$ and Lemma~\ref{t:526}
with $\eta$ replaced by $\frac 23 \eta $, for any $\epsilon$
small enough we have
\begin{eqnarray*}
&& \nu_\epsilon\Big( \exists\, t\in [\sigma\ell,\ell] \, : \:
x(t) \notin \big(\bam_{a_\ell}(t)-\epsilon^{\frac 12 - \frac 23 \eta},
1+{\textstyle\frac 12} \epsilon^{\frac 12 - \eta}\big) \Big) 
\\ && \qquad \qquad \qquad \le \exp\big\{-\epsilon^{-\frac 23 \eta}
\big\} + \exp\big\{-\epsilon^{-\eta}\big\} + \exp\big\{-e^{\frac 14
a_\ell}\big\}.
\end{eqnarray*}
By choosing $\sigma\in [\sigma_0,1)$ with $\sigma_0$ as before, 
the previous estimate implies
\begin{eqnarray*}
&& \nu_\epsilon\Big(\sup_{t\in[\sigma\ell,\ell]} \big| x(t) -
\bam_{\mc Z(x)}(t) \big| > \epsilon^{\frac 12-\eta}, |\mc Z(x)| \le 
a_\ell \Big)
\nn \\ && \phantom{mmmmm} \le \exp\big\{-\epsilon^{-\eta}\big\} +
\exp\big\{-\epsilon^{-\frac 23\eta}\big\} + \exp\big\{-e^{\frac 14 
a_\ell}\big\}.
\end{eqnarray*}
Finally, by applying Lemma~\ref{t:53} with $\eta$ replaced by 
$\frac 23\eta$, we have that, for any $\sigma' \in 
\big(0, 1-\frac 43\eta\big)$ and $\epsilon$ small enough (which implies
$a_\ell \le\sqrt\ell$),
\begin{eqnarray*}
&& \nu_\epsilon\Big(\sup_{t\in[-\sigma'\ell+a_\ell, 
\sigma'\ell -a_\ell]}
\big| x(t) - \bam_{\mc Z(x)}(t) \big| > \epsilon^{\frac 12-\eta},
|\mc Z(x)| \le a_\ell \Big) \nn \\ && \phantom{mmmm}\le
\nu_\epsilon\Big(\sup_{t\in[-\sigma'\ell+a_\ell, \sigma'\ell -a_\ell]}
\big| x(t) - \bam_{\mc Z(x)}(t) \big| > \epsilon^{\frac 12 - 
\frac 23\eta}, |\mc Z(x)| \le a_\ell \Big) \nn \\ && \phantom{mmmm} \le
\exp\big\{-\epsilon^{-\frac 23 \eta}\big\}.
\end{eqnarray*}
Since $\sigma_0 < 1-\frac 43\eta$, we can choose 
$\sigma\in (\sigma_0,1)$ and
$\sigma'\in(\sigma,1)$; the bound \eqref{pa2} follows.
\qed

\section{Weak convergence of the measure}
\label{sec:67}

We first show, by using the representation of the measure $\mu_\epsilon$
given in Proposition~\ref{t:61} and the sharp estimates of the previous section, that $\mu_\epsilon$ concentrates in a 
$\sqrt\epsilon$-neighborhood of the manifold $\mc M$. We also show 
that the interface remains in a compact set of $\bb R$ with 
probability close to one. Recall that $\mc Z(x)$ is the leftmost zero
of $x\in \mc X_\ell$.

\begin{theorem}
\label{t:unifico}
For each $\eta>0$ we have
\begin{eqnarray}
\label{te}
    && \lim_{\epsilon\to 0} \mu_\epsilon 
    \big( \big\{x\, : \: d(x,\mc M) > \epsilon^{\frac 12 -\eta}\big\} \big) = 0,
\\ \label{nuovo} &&
\lim_{L\to \infty}\varlimsup_{\epsilon\to 0} \mu_\epsilon
\big( \big\{x\, :\, |\mc Z(x) | > L\big\}\big) = 0.
\end{eqnarray}
   \end{theorem}

We first prove a rougher bound showing that, uniformly in $\epsilon$,
the measure $\mu_\epsilon$ of bounded sets is close to one. 

\begin{lemma}
\label{bbs}
We have
\begin{equation*}
\lim_{K\to\infty}\; \varlimsup_{\epsilon\to 0}\, \mu_\epsilon
\big( \big\{x\, : \: \|x\|_\infty > K \big\} \big) = 0.
\end{equation*}
\end{lemma}

\proof We assume that $K\in\bb N$ ad use the representation of 
$\mu_\epsilon$ in Proposition~\ref{t:61} together with the bound
$|\partial_{xx}S_0(t,x)| \le A_3(1+|x|)$ proven in Theorem~\ref{t:31}. 
We have 
\begin{eqnarray*}
\mu_\epsilon \big( \big\{x\, : \: \|x\|_\infty > K \big\} \big) & = & 
\sum_{h\ge K} \frac{\nu_\epsilon\Big(e^{-\frac 12\int_{-\ell}^\ell
\!dt\, \partial_{xx}S_0(\ell-t,x(t))}\id_{\{\|x\|_\infty 
\in[h,h+1)\}} \Big)}
{\nu_\epsilon \Big(e^{-\frac 12\int_{-\ell}^\ell 
\!dt\, \partial_{xx}S_0(\ell-t,x(t))}\Big)} \\
& \le & \sum_{h\ge K} \frac{e^{A_3\ell(2+h)} \,
\nu_\epsilon\big( \|x\|_\infty\ge h\big)}
{e^{-3A_3\ell}\,\nu_\epsilon\big( \|x\|_\infty\le 2\big)} 
\\ &\le & \sum_{h\ge K} 2\, e^{A_3\ell(5+h)}\, 
e^{-c_0 (\epsilon\ell)^{-1}h^2},
\end{eqnarray*}
where we used that, by Lemma~\ref{t:52}, if
$\epsilon$ is small enough then 
$\nu_\epsilon\big( \|x\|_\infty\le 2\big) 
\ge \frac 12$ and $\nu_\epsilon\big( \|x\|_\infty\ge h\big) \le 
e^{-c_0 (\epsilon\ell)^{-1}h^2}$ for some $c_0>0$. 
\qed

\medskip\noindent
{\it Proof of Theorem~\ref{t:unifico}.} 
We first prove \eqref{te}. By Lemma \ref{bbs} it is enough to show 
that for each $K\ge 2$ and $\eta>0$ we have
   \begin{equation}
   \label{enough}
\lim_{\epsilon\to 0} \mu_\epsilon 
\big(\mc B \big) = 0, \qquad
\mc B := \big\{x\, : \: d(x,\mc M) > \epsilon^{\frac 12 -\eta} \big\} \cap \big\{x\, : \:\|x\|_\infty \le K\big\},
   \end{equation}
By the representation given in Proposition~\ref{t:61} and 
Theorem~\ref{t:31}, 
   \begin{eqnarray*}
\mu_\epsilon\big(\mc B\big) & = &
\frac{\nu_\epsilon\bigg( \exp\Big\{ - \frac 12 
\int_{-\ell}^\ell\!dt\,\partial_{xx} S_0(\ell-t,x(t))\Big\}
\id_{\mc B}\bigg)}
{ \nu_\epsilon\bigg( \exp\Big\{ - \frac 12 
\int_{-\ell}^\ell\!dt\,\partial_{xx} S_0(\ell-t,x(t))\Big\}\bigg)}
\\ & \le & e^{A_3(4+K) \ell} \,\frac{\nu_\epsilon\big(d(x,\mc 
M) > \epsilon^{\frac 12 -\eta} \big)}
{\nu_\epsilon\big(\|x\|_\infty\le 2\big)},
   \end{eqnarray*}
which, by \eqref{5.8}, concludes the proof of \eqref{te}.

We next prove \eqref{nuovo}. By \eqref{te} it is enough to show 
that, for some $\eta>0$, 
   \begin{equation}
   \label{gl}
\lim_{L\to \infty}\varlimsup_{\epsilon\to 0} \mu_\epsilon
\big( \mc B_{\epsilon,L} \big) = 0, \qquad \mc B_{\epsilon,L} :=
\big\{x\, :\: |\mc Z(x) | > L,\; d(x,\mc M) \le \epsilon^{\frac 12 -\eta} \big\} .
   \end{equation}
Let
\begin{equation*}
\mc I_\ell(x) := - \frac 12 
\int_{-\ell}^\ell\!dt\,\partial_{xx} S_0(\ell-t,x(t)) - \frac 12 
\log\ell. 
\end{equation*}
By the representation given in Proposition~\ref{t:61}, 
\begin{equation*}
\mu_\epsilon \big( \mc B_{\epsilon,L} \big) = 
\frac{\nu_\epsilon\big(e^{\mc I_\ell} \, \id_{\mc B_{\epsilon,L} } 
\big)}{\nu_\epsilon\big( e^{\mc I_\ell} \big)}.
\end{equation*}
We first observe that, by setting
\begin{equation*}
\mc A := \big\{d(x,\mc M) 
\le \epsilon^{\frac 12 - \eta} \big\} \cap\{ |\mc Z(x)| \le a\} \cap 
\big\{x(t)\ge \bam_a(t) \,\, \forall\, t\in [-\ell+a,a]\big\},
\end{equation*}
we have
\begin{equation*}
\nu_\epsilon\big( e^{\mc I_\ell} \big) \ge 
\nu_\epsilon\big( e^{\mc I_\ell} \, \id_{\mc A} \big)
\ge \, e^{-A_5 a}\, \big(1-\nu_\epsilon(\mc A^\complement)\big),
\end{equation*}
where we used Proposition~\ref{t:32b}. By choosing $a$ large enough and
applying Theorem~\ref{t:ap} we get $\varliminf_{\epsilon\to 0}
\nu_\epsilon\big(e^{\mc I_\ell}\big)  >0$.

We next observe that, by Theorem~\ref{t:31}, we have 
$|\partial_{xx} S_0(\ell-t,x(t))| \le 3 A_3$ on the event 
$d(x,\mc M) \le 1$, so that, for $\ell$ large enough, 
\begin{equation*}
\mc B_{\epsilon,L} = \Big( \mc B_{\epsilon,L} \cap
\big\{|\mc I_\ell| \le 2A_5 L \big\} \Big)
\cup 
\bigcup_{h=L}^{\big[7 \,\frac{A_3}{A_5}\, \ell\big]} 
\Big(  \mc B_{\epsilon,L} \cap \big\{2hA_5 < |\mc I_\ell| 
\le 2A_5 (h+1) \big\} \Big).
\end{equation*}
Accordingly,
\begin{equation*}
\nu_\epsilon\big(
 e^{\mc I_\ell} \, \id_{\mc B_{\epsilon,L} } \big)
\, \le  \,
e^{2A_5 L } \nu_\epsilon\big(|\mc Z | > L \big) 
+ \sum_{h=L}^{\big[7\,\frac{A_3}{A_5}\, \ell\big]}\, e^{2 A_5 (h+1) } 
\, \nu_\epsilon
\Big( \mc B_{\epsilon,L} \cap \big\{|\mc I_\ell| > 2h A_5  \big\}\Big).
\end{equation*}
Choosing $\eta$ small enough and applying Proposition \ref{t:32b} 
we get
\begin{eqnarray*}
&& \nu_\epsilon \Big( \mc B_{\epsilon,L} \cap 
\big\{|\mc I_\ell| > 2hA_5  \big\}\Big) \, \le \, 
\nu_\epsilon\big( d(x,\mc M) \le \epsilon^{\frac 12 - \eta},
\, |\mc I_\ell| > 2A_5 \, h \big)
\\ && \qquad \qquad\qquad\le  \nu_\epsilon\big(|\mc Z(x)|>h\big) 
\, + \, \nu_\epsilon\big(\{\exists\,t\in[-\ell+h,h]\, : \,
x(t)<\bam_h(t)\}\big)
\end{eqnarray*}
By choosing $L$ large enough and applying Theorem \ref{t:ap} we thus obtain
\begin{equation*}
\varlimsup_{\epsilon\to 0} \nu_\epsilon\big(
e^{\mc I_\ell} \, \id_{\mc B_{\epsilon,L} } \big) \le
2\exp\big\{2A_5L - e^{\frac 14 L}\big\} + \sum_{h=L}^\infty
3 \exp\big\{2 A_5 (h+1) - e^{\frac 14 h}\big\},
\end{equation*}
which concludes the proof.
\qed

\bigskip
We next conclude the proof of Theorem~\ref{t:1} by characterizing
the limit points of $\mu_\epsilon$ as the invariant measure of
\eqref{feq-2}.    
By \cite[Thm.\ 5.1]{FV} $\mu_\epsilon$ is the unique invariant measure
of the process $X=X_\sigma$ in $C(\bb R_+;\mc X_\ell)$ which solves 
\eqref{feq-1} with $\ell=\frac 14\log\epsilon^{-1}$. For $T>0$,
we denote by $\bb P^\epsilon_{x_0}$ the law of the process 
$X_{\epsilon^{-1}\sigma}$, $\sigma\in [0,T]$, where $X$ is the 
solution to \eqref{feq-1} with initial datum $x_0\in\mc X_\ell$. 
We regard $\bb P^\epsilon_{x_0}$ as a probability on $C([0,T];\mc X)$,
endowed with the topology of uniform convergence. Let also $P_{z_0}$ 
be the law of the one-dimensional diffusion solution to \eqref{feq-2} 
with initial datum $z_0\in\bb R$. We finally define $\bb P_{z_0}$ 
as the probability measure on $C([0,T];\mc X)$ with support
$C([0,T];\mc M)$ such that  $\bb P_{z_0} (A) = P_{z_0}
\big(\bam_{\zeta_\cdot}\in A\big)$. The analysis in \cite{bbbdy}, 
see in particular Theorem~2.2, yields the weak convergence of 
$\bb P^\epsilon_{x_0}$ to $\bb P_{\mc Z(x_0)}$, recall $\mc Z(x)$ is 
the leftmost zero of $x$. Moreover, for $\eta$ small enough, the 
above convergence is uniform for $\mc Z(x_0)$ in compacts and $x_0$ 
such that $d(x_0,\mc M)\le \epsilon^{\frac 12 - \eta}$. 

\begin{theorem}
\label{t:10}
Let $T>0$. There exists $\eta_1>0$ such that for any $\eta\in
[0,\eta_1]$ the following  holds. For each $L>0$ and each 
uniformly continuos and bounded function $F$ on $C([0,T];\mc X)$ 
we have
\begin{equation}
  \label{eq:1.3}
\lim_{\epsilon\to 0} \:\: \sup_{z_0\in [-L,L]} \:\:
\sup_{x_0\in \mc N^\epsilon_\eta(z_0)} \:\: \big| \bb P^\epsilon_{x_0}
(F) - \bb P_{z_0}(F) \big| = 0,
\end{equation}
where $\mc N^\epsilon_\eta(z_0) := \big\{x \in \mc X_\ell \, :\,
d(x,\bam_{z_0}) \le \epsilon^{\frac 12 - \eta} \big\}$.
\end{theorem}

\noindent\textit{Proof of Theorem~\ref{t:1}.}
We set $p_\epsilon := \mu_\epsilon \circ \mc Z^{-1}$, namely
$p_\epsilon$ is the distribution of the real random variable
$\mc Z(x)$ when $x$ is distributed according to $\mu_\epsilon$. 
Note that $p_\epsilon$ is tight by \eqref{nuovo}. Let also
$Q_z(\cdot)$ be a regular version of the conditional probability
$\mu_\epsilon\big(\,\cdot\, | \, \mc Z=z\big)$. 

Denote by $\pi_\sigma : C(\bb R_+;\mc X) \to \mc X$ the evaluation map
at $\sigma$. Since $\mu_\epsilon$ is the invariant measure 
of \eqref{feq-1}, for each $\sigma\in\bb R_+$ and each uniformly
continuous and bounded function $F$ on $\mc X$, we have 
\begin{eqnarray}
  \label{pippo}
\int\!d\mu_\epsilon(x) \, F(x) & = & \int\!d\mu_\epsilon(x_0) \, 
\bb P^\epsilon_{x_0} \big(F \circ \pi_\sigma\big) 
\nn \\ & = &  \int_{-L}^L\!dp_\epsilon(z_0) 
\int_{\mc N^\epsilon_\eta(z_0)}\!dQ_{z_0}
(x_0)\, \bb P^\epsilon_{x_0} \big(F \circ \pi_\sigma\big)
+ R_{L,\epsilon}(F), \qquad
\end{eqnarray}
where, by the compactness and tube estimate of Theorem~\ref{t:unifico},
for each $\eta>0$ we have 
\begin{eqnarray*}
&& \!\!\! \!\!\! \!\!\! \!\!\! 
\lim_{L\to\infty} \: \varlimsup_{\epsilon\to0} 
\big|R_{L,\epsilon}(F) \big| \\ && \quad \le \|F\|_\infty \: 
\lim_{L\to\infty} \: \varlimsup_{\epsilon\to 0} \mu_\epsilon
\Big(\big\{x\, : \, \big|\mc Z(x) \big|>L\big\} \cup
\big\{x\, :\, d(x,\bam_{\mc Z(x)})>\epsilon^{\frac 12 -\eta}\big\}
\Big) = 0.
\end{eqnarray*}

By the tightness of $p_\epsilon$, there exists a probability measure 
$p$ on the $\bb R$ and a subsequence, still denoted by $p_\epsilon$,
weakly convergent to $p$. By Theorem~\ref{t:10}, for any 
$\eta\in [0,\eta_1]$, the real function 
$$
z_0 \: \mapsto \: \int_{N^\epsilon_\eta(z_0)}\!dQ_{z_0}
(x_0)\, \bb P^\epsilon_{x_0} \big(F \circ \pi_\sigma\big)
$$
converges to $\bb P_{z_0} \big(F \circ\pi_\sigma \big) =
P_{z_0}\big(F(\bam_{\zeta_\sigma}\big)\big)$ uniformly for $z_0$ in
compacts. By taking in \eqref{pippo} the limit $\epsilon\to 0$  
along the converging subsequence and then $L\to \infty$, 
we get 
\begin{equation}
\label{pippo2}
\lim_{\epsilon\to 0}  
\int\!d\mu_\epsilon(x) \, F(x)  = \int\!d p (z_0)\, 
P_{z_0} \big( F\big(\bam_{\zeta_\sigma} \big)\big).
\end{equation}
By the arbitrariness of $\sigma$ and $F$, \eqref{pippo2}
shows that $p$ is an invariant measure for the one-dimensional 
diffusion process \eqref{feq-2}. Since the latter has a unique
invariant measure given by $\hat\mu$ as in equation \eqref{hatmu},
we conclude that $p=\hat \mu$, and, by \eqref{pippo2}, the proof of the
theorem.  
\qed

\medskip\noindent{\it Remark.}
From the above proof it follows that the stationary process 
associated to \eqref{feq-1} (as a random element in $C(\bb R;\mc X)$)
converges in law to $\bam_{\zeta_\cdot}$, where $\zeta_\cdot$
is the stationary process associated to \eqref{eq:1.3}.

\appendix
\section{Weierstrass analysis of the mechanical problem}
\label{s:ap}

Recall that $S$ has been defined in \eqref{S=}. Since $x=1$ is a global
maximum of the potential $-2V$, for each $(t,x)\in (0,\infty)\times
\bb R$, there is a unique solution $\psi_{t,x}(\cdot)$ to the Newton 
equation 
    \begin{equation}
   \label{eleq}
 \begin{cases}
 \ddot\psi_{t,x}=2V'(\psi_{t,x})\\
 \psi_{t,x}(0)=1\\
 \psi_{t,x}(t)=x.
 \end{cases}
 \end{equation}
As discussed in the proof  of Proposition \ref{t:41},  
$\psi_{t,x}(\cdot)$ is the minimizer for $S(t,x)$, that is, 
 \begin{equation}
   \label{defS}
   S(t,x)=\int_0^t\!ds\, \Big[
   \frac 12 \dot\psi_{t,x}(s)^2 + 2V(\psi_{t,x}(s))\Big].
   \end{equation}
Integration of \eqref{eleq} yields that, for $s\in(0,t)$, 
\begin{equation}
\label{eqi}
\dot\psi_{t,x}(s)^2 - 4 V(\psi_{t,x}(s))=e_{t,x},
\end{equation}
for some non-negative constant $e_{t,x}$. Clearly $e_{t,1} = 0$;
otherwise, integrating \eqref{eqi} by separation of variables, we get 
that $e_{t,x}$ solves
\begin{equation}
\label{defe}
t = \bigg\vert\int_x^1\!  \frac{du}{\sqrt{4V(u)+e_{t,x}}}\bigg\vert.
\end{equation}
Also, substitution of \eqref{eqi} into \eqref{defS} gives 
   \begin{equation} 
   \label{forS}
S(t,x)= \bigg\vert \int_x^1\!  du\,\sqrt{4V(u)+e_{t,x}}\bigg\vert
\,-\,\frac{1}{2}\,t\,e_{t,x}.
\end{equation}
We finally notice that, by the symmetry of $V$, $\inf_{\mc X_\ell}
\mc F_\ell= S(2\ell,-1)$. 

In the first two lemmata we prove the estimates used in 
Section~\ref{sec:4} to prove the variational convergence of 
$\mc G_\ell$.

    \begin{lemma}
   \label{t:33}
Let $E_\ell := e_{2\ell,-1}$. Then
\begin{eqnarray}
   \label{El}
&& \lim_{\ell\to \infty} e^{4\ell} \, E_\ell = 64,
\\ && \label{sigma*}
\lim_{\ell\to\infty} e^{4\ell}\Big[S(2\ell,-1) - \frac 43\Big]=16.
\end{eqnarray}   
Moreover, there exists a constant $A_1>$ such that for any $\ell\ge 1$,
  \begin{equation}
   \label{bm0i}
\sup_{t\in [-\ell,\ell]}\big| m^*_\ell(t)-\bam_0(t) \big| 
\le A_1 \, e^{-2\ell}.
   \end{equation}
   \end{lemma}
 
\noindent\emph{Proof.} 
Direct integration yields 
\begin{equation}
\label{int}
\bigg|\int_x^1\!\frac{du}{\sqrt{\gamma^2(1-u)^2 + \beta}}\bigg|
= \frac 1\gamma \avarsh \frac{\gamma|x-1|}{\sqrt{\beta}}, \qquad
\beta,\gamma>0,
\end{equation}
which will be repeatedly used in the sequel. 
By \eqref{defe} and the symmetry of $V$ we thus have
\begin{equation*}
\ell = \int_0^1\! \frac{du}{\sqrt{4V(u)+E_\ell}} = 
\frac 12 \avarsh \frac {2}{\sqrt {E_\ell}} + R_1(E_\ell)
\end{equation*}
where 
\begin{equation}
\label{R1}
R_1(E) : = \int_0^1\! \frac{du}{\sqrt{4V(u)+E}} -
\int_0^1\! \frac{du}{\sqrt{4(1-u)^2+E}}.
\end{equation}
A straightforward computation yields 
\begin{equation}
\label{a.9.5}
\lim_{E\downarrow 0} R_1(E) = \frac 12 
\int_0^1\, \frac{du}{1+u} = \frac 12\log 2.
\end{equation}
Since $E_\ell\downarrow 0$ as $\ell\to +\infty$, \eqref{El}
follows. 

To prove \eqref{sigma*} we first observe that 
$\frac 43 = \int_{-1}^1\!du\, \sqrt{4V(u)}$. We next show that
\begin{equation}
\label{p100}
\lim_{\ell\to\infty} \frac{1}{E_\ell} \Big[S(2\ell,-1) - 
\int_{-1}^1\!du\, \sqrt{4V(u)}\Big] = \frac 14,
\end{equation}
which, together with \eqref{El}, yields \eqref{sigma*}. The
identities \eqref{defe}, \eqref{forS}, and simple computations give
\begin{equation*}
\frac{1}{E_\ell} \Big[S(2\ell,-1) - 
\int_{-1}^1\!du\, \sqrt{4V(u)}\Big] = R_2(E_\ell),
\end{equation*}
where
\begin{equation*}
R_2(E) :=  \int_0^1\!du\, \frac{E}{\sqrt{4V(u)+E}\,
\Big(\sqrt{4V(u)}+ \sqrt{4V(u)+E}\Big)^2}.
\end{equation*}
By the change of variable $1-u=\frac 12 \sqrt{E}y$ we get
$$
R_2(E) = \frac 12 \int_0^{\frac 2{\sqrt E}}\!\frac{dy}{\sqrt{1+
y^2(1-\frac 14 \sqrt E y)^2} \Big(y(1-\frac 14 \sqrt E y)+ \sqrt{1+
y^2(1-\frac 14 \sqrt E y)^2}\Big)^2},
$$
hence
$$
\lim_{E\downarrow 0} R_2(E) = \frac 12 \int_0^{\infty}\!
\frac{dy}{\sqrt{1+y^2}\, \Big(y+ \sqrt{1+y^2}\Big)^2} = \frac 14,
$$
which gives \eqref{p100}.

To prove \eqref{bm0i} we first note that it is enough to consider 
the case $t\ge 0$ as both $m^*_\ell$ and $\bam_0$ are odd functions.
Since $m^*_\ell$ is the solution to \eqref{ml}, for $t\in [0,\ell]$, 
we have $e_{\ell-t,m^*_\ell(t)} = e_{2\ell,-1} = E_\ell$, namely
\begin{equation*}
t\,=\, \int_0^{m^*_{\ell}(t)}\! \frac{du}{\sqrt{4V(u) + E_{\ell}}}.
\end{equation*}
On the other hand, since $\bam_0'=\sqrt{4V(\bam_0)}$, for $t\ge 0$,
\begin{equation*}
t \, = \, \int_0^{\bam_0(t)}\! \frac{du}{\sqrt{4V(u)}}.
\end{equation*}
Then,
\begin{equation}
\label{m1-mo}
\int ^{m^*_\ell(t)}_{\bam_0(t)} \!\frac{du}{\sqrt{4V(u) + E_{\ell}}} 
\, = \, \int_0^{\bam_0(t)}\! du \bigg[ \frac{1}{\sqrt{4V(u)}} -
\frac{1}{\sqrt{4V(u)+ E_{\ell}}}\bigg].
\end{equation}
We now have
\begin{equation*}
\frac{1}{\sqrt{4V(u)}} - \frac{1}{\sqrt{4V(u)+ E_{\ell}}} \le
\frac{E_\ell}{2\big[4V(u)\big]^{\frac 32}}
\end{equation*} 
and 
\begin{equation*}
\int ^{m^*_\ell(t)}_{\bam_0(t)} \!\frac{du}{\sqrt{4V(u) + E_{\ell}}}
\, \ge \,\frac{\big(m^*_\ell(t)-\bam_0(t)\big)}{\sqrt{4V(\bam_0(t)) 
+ E_{\ell}}}\, = \, 
\frac{\big(m^*_\ell(t)-\bam_0(t)\big)}{\sqrt{\bam_0'(t)^2 + E_{\ell}}}.
\end{equation*}
After substituting in \eqref{m1-mo}, we obtain
\begin{eqnarray*}
m^*_\ell (t)-\bam_0(t) & \le & 
E_{\ell}\,\sqrt{\bam_0'(t)^2  + E_{\ell}} \,
\int^{\bam_0(t)}_0 \frac{du}{2\big[4V(u)\big]^{\frac 32}} 
\\ & = & E_{\ell}\,\sqrt{\bam_0'(t)^2  + E_{\ell}} \,
\int^{t}_0 \frac{ds}{2\,\bam_0'(s)^2}, 
\end{eqnarray*}
that, recalling \eqref{El}, yields \eqref{bm0i}
\qed

    \begin{lemma}
   \label{t:A4}
Let $x_\ell$ be a sequence in $\mc X$ for which there exists
a constant $C>0$ such that  
$\mc F_\ell(x_\ell) - \mc F_\ell(m^*_\ell) \le C\,e^{-4\ell}$ for any
$\ell\ge 1$. Then $x_\ell$ has a converging subsequence. 
   \end{lemma}
 
\proof Pre-compactness of a sequence $x_\ell$ in $\mc X$ is equivalent
to its equi-continuity together with 
\begin{equation}
\label{compac}
\lim_{K\to \infty}\: \varlimsup_{\ell\to\infty}\: 
\sup_{t\in \pm  [K,\infty)} \; \big|x_\ell(t)\mp 1\big| = 0.
\end{equation}

Pick a sequence $x_\ell$ in $\mc X$ for which there exists
a constant $C>0$ such that $\mc G_\ell(x_\ell)\le C$. Equivalently,
by \eqref{stima**}, there exists $C>0$ such that $\mc F_\ell(x_\ell) 
\le \frac 43 + C \,e^{-4\ell}$. This estimate implies immediately the
equi-continuity of the sequence $x_\ell$. 

We next prove that 
\begin{equation}
\label{comp1}
\varliminf_{K\to \infty} \: \varliminf_{\ell\to \infty} \:
\inf_{t\in  (-\infty, -K]} \; x_\ell(t) \ge -1.
\end{equation}
Given $\delta>0$, let $\tau_\ell^\delta := \inf \, \{ t\in [-\ell,\ell]
\, : \, x_\ell(t) = -1-\delta\} \wedge \ell$ be the time of the first
passage by $-1-\delta$. The estimate \eqref{comp1} is then equivalent 
to $\varliminf_\ell \tau_\ell^\delta> - \infty$ for any $\delta>0$.
Since $x_\ell\in\mc X_\ell$ we have
$\tau_\ell^\delta \in (-\ell,\ell]$. If $\tau_\ell^\delta = \ell$
\eqref{comp1} holds trivially, otherwise we define 
$\sigma_\ell^\delta := \sup \Big\{t\in [-\ell,\tau_\ell^\delta] 
\, : \, x_\ell(t) = -1-\frac \delta2 \Big\}$. Recalling the notation
\eqref{Fab}, the equi-boundedness of the excess free energy 
$\mc G_\ell(x_\ell)$ then yields
\begin{eqnarray*}
C \, e^{-4\ell} & \ge & \mc F_\ell(x_\ell) - \mc F_\ell(m^*_\ell)
\\ & = & \mc F_{[-\ell,\tau_\ell^\delta]}(x_\ell) - 
\mc F_{[-\ell,\tau_\ell^\delta]}(m^*_\ell) + 
\mc F_{[\tau_\ell^\delta,\ell]}(x_\ell) - 
\mc F_{[\tau_\ell^\delta,\ell]}(m^*_\ell)
\\ & \ge & \int_{\sigma_\ell^\delta}^{\tau_\ell^\delta}\!dt\,
2 V(x_\ell(t)) - \mc F_{[-\ell,\tau_\ell^\delta]}(m^*_\ell)
\\ && + \inf_{\substack{x\in\mc X_\ell \\ 
x(\tau_\ell^\delta) = -1- \delta}}
\mc F_{[\tau_\ell^\delta,\ell]}(x) - \mc F_{[\tau_\ell^\delta,\ell]}
(m^*_\ell).
\end{eqnarray*}
Recalling \eqref{defS}, the second difference on the r.h.s.\ above
equals $S(\ell-\tau_\ell^\delta,-1-\delta) -
S(\ell-\tau_\ell^\delta,m^*_\ell(\tau_\ell^\delta))$. Since 
$x\mapsto S(t,x)$ is increasing and $m^*_\ell(\tau_\ell^\delta)> - 1$
we conclude that
\begin{equation*}
C\, e^{-4\ell} \ge 2 V\Big(-1-\frac\delta2\Big) \,
\big(\tau_\ell^\delta - \sigma_\ell^\delta\big) - 
\mc F_{[-\ell,\tau_\ell^\delta]} (m^*_\ell),
\end{equation*}
whence
\begin{equation*}
\varliminf_{\ell\to\infty} \, \mc F_{[-\ell,\tau_\ell^\delta]} 
(m^*_\ell) \, \ge \,
\varliminf_{\ell\to\infty} \, 2 V\Big(-1-\frac\delta2\Big) \,
\big(\tau_\ell^\delta - \sigma_\ell^\delta\big) >0,
\end{equation*}
where we used the equi-continuity of $x_\ell$. We then conclude
that $\varliminf_\ell \tau_\ell^\delta> - \infty$ and \eqref{comp1}
follows since $\delta>0$ was arbitrary. By symmetry we also
have
\begin{equation}
\label{comp2}
\varlimsup_{K\to \infty} \: \varlimsup_{\ell\to \infty} \:
\sup_{t\in  [K,\infty)} \; x_\ell(t) \le 1.
\end{equation}

We next prove
\begin{equation}
\label{comp3}
\varlimsup_{K\to \infty} \: \varlimsup_{\ell\to \infty} \:
\sup_{t\in  (-\infty,- K]} \; x_\ell(t) \le -1.
\end{equation}
Given $\delta\in(0,1)$ let 
$T\equiv T_\ell^\delta:= 
\inf\{t\in (-\ell,\ell)\, :\, x_\ell(t) = -1+\delta\}$; by
continuity of $x_\ell$ it follows that $T\in (-\ell,\ell)$. We have
\begin{eqnarray*}
C \, e^{-4\ell} & \ge & \mc F_\ell(x_\ell) \, - \, \frac 43 
\, \ge \, \inf_{\substack{x\in\mc X_\ell \\ x(T) = -1+ \delta}}
\mc F_{[-\ell,T]}(x) \, + \, \inf_{\substack{x\in\mc X_\ell \\ x(T) = -1+ \delta}}
\mc F_{[T,\ell]}(x) \, - \, \frac 43
\\ & = & S(\ell+T,1-\delta) \, + \, S(\ell-T,-1+\delta) \, - \, \frac 43
\\ & = & \int_{1-\delta}^1\!du\, \Big[\sqrt{4V(u)+E_+} - \sqrt{4V(u)}\Big] 
\, - \, \frac 12 (\ell+T) \, E_+  
\\ && +\, \int_{-1+\delta}^1\!du\, \Big[\sqrt{4V(u)+E_-} - \sqrt{4V(u)}\Big] 
\, - \, \frac 12 (\ell - T) \, E_-,
\end{eqnarray*}
where $E_\pm>0$ is the solution to
\begin{equation}
  \label{comp31}
  \ell \pm T = \int_{\pm (1-\delta)}^1\! \frac{du}{\sqrt{4V(u)+E_\pm}}
\end{equation}
and we used the symmetry of $V$, the identity \eqref{forS}, and
$\frac 43 = \int_{-1}^1\!du\,\sqrt{4V(u)}$. 

By computations similar to those used in proving \eqref{sigma*} we get
\begin{equation*}
\frac{E_\pm^2}2 \int_{\pm(1-\delta)}^1\! \frac{du}{\sqrt{4V(u)+E_\pm}\,
\Big(\sqrt{4V(u)}+ \sqrt{4V(u)+E_\pm}\Big)^2} \le C \, e^{-4\ell},
\end{equation*}
which gives that for each $\delta\in (0,1)$ we have
\begin{equation}
\label{precomp}
\varlimsup_{\ell\to\infty} \, e^{4\ell} \, E_\pm  < \infty. 
\end{equation}
We rewrite \eqref{comp31} for $E_+$ as 
\begin{eqnarray*}
  \ell + T & = & \int_0^1\! \frac{du}{\sqrt{4V(u)+E_+}} -
  \int_0^{1-\delta} \! \frac{du}{\sqrt{4V(u)+E_+}} 
  \\
  &= &
  \frac 12 \avarsh \frac {2}{\sqrt {E_+}} + R_1(E_+) 
  -\int_0^{1-\delta} \! \frac{du}{\sqrt{4V(u)+E_+}},
\end{eqnarray*}
where we recall that $R_1$ is defined in \eqref{R1}. 
By taking the limit $\ell\to\infty$ in this identity and using
the estimate \eqref{precomp} for $E_+$ together with \eqref{a.9.5},
we get that, for each $\delta\in(0,1)$, 
\begin{equation*}
  \varliminf_{\ell\to\infty} T_\ell^\delta 
   \ge  
  \varliminf_{\ell\to\infty} 
  \Big[ \frac 12 \avarsh \frac {2}{\sqrt {E_+}}-\ell \Big]
  + \frac 12\log 2 - \avarth(1-\delta) > -\infty,
\end{equation*}
which yields \eqref{comp3}. By symmetry
\begin{equation}
\label{comp4}
\varliminf_{K\to \infty} \: \varliminf_{\ell\to \infty} \:
\inf_{t\in  [K,\infty)} \; x_\ell(t) \ge 1.
\end{equation}
The estimate \eqref{compac} follows from \eqref{comp1}, \eqref{comp2}, 
\eqref{comp3}, and \eqref{comp4}.
\qed

\bigskip
Recall that $S_0$ has been defined in \eqref{S0=}. The regularity of 
$S$, whence of $S_0$, is standard for $t>0$ and $x\in \bb R$. In the 
next theorem, we show that $S_0$ is actually regular for 
$t\downarrow0$ and estimate its second derivative. 

    \begin{theorem}
   \label{t:31}
We have
$$
\lim_{t\downarrow 0} S_0(t,x) \, = \,
\lim_{t\downarrow 0}\partial_x S_0(t,x)  \, = \, 
\lim_{t\downarrow 0}\partial_{xx} S_0(t,x) \, = \, 0,
$$ 
uniformly for $x$ in compacts. Moreover
\begin{equation}
\label{bddS}
A_3 := \sup_{(t,x) \in \bb R_+\times \bb R} \frac{1}{1+|x|} \, 
\big|\partial_{xx} S_0(t,x)\big|  < + \infty.
\end{equation}
   \end{theorem}
   
\proof 
Set $c(x) := 2+|x|$. Then  $0\le 4V(u) \le c(x)^2(1-u)^2$ 
whenever $x<1$ and $u\in [x,1]$ or $x>1$ and $u\in [1,x]$.
Recalling \eqref{int}, from \eqref{defe} we get
\begin{equation*}
\frac{1}{c(x)}\avarsh\frac{c(x) \, |1-x|}{\sqrt{e_{t,x}}}
\le t \le \frac{|1-x|}{\sqrt{e_{t,x}}},
\end{equation*}
whence
\begin{equation}
\label{eg}
\sqrt{e_{t,x}} = \frac{|1-x|}t \, \big[1+g(t,x)\big] \quad
\text{ with }\quad
\frac{c(x)\, t}{\varsh[c(x)\,t]}-1 \le g(t,x)\le 0.
\end{equation}
Then, setting $A(V,e) := \sqrt{4V+e}-\sqrt e$ and recalling
\eqref{S0=}, \eqref{forS}, 
\begin{eqnarray*}
S_0(t,x) & = & |1-x| \, \sqrt{e_{t,x}} + \bigg\vert\int_x^1\!du\,  
A(V(u),e_{t,x}) \bigg\vert  \,-\,\frac{1}{2}\,t\,e_{t,x} - 
\frac{(1-x)^2}{2t} \\ & = & \bigg\vert\int_x^1\!du\,  
A(V(u),e_{t,x}) \bigg\vert - \frac{(1-x)^2}{2t} g(t,x)^2.
\end{eqnarray*}
Analogously, recalling \eqref{b=}, 
\begin{eqnarray*}
\partial _x S_0(t,x) & = & \sgn(x-1) \sqrt{4V(x)+e_{t,x}}
\,-\,\frac{x-1}{t} 
\\ & = & \sgn(x-1) A(V(x),e_{t,x}) - \frac{1-x}{t}g(t,x).
\end{eqnarray*}
Since $A(V,e) \le \frac{2V}{\sqrt{e}}$, we now have
\begin{equation*}
|S_0(t,x)| \le \frac{t}{1+g(t,x)}\, \frac{2}{1-x}\, 
\int_x^1\!du\, V(u) \, + \, \frac{(1-x)^2}{2t} g(t,x)^2
\end{equation*}
and 
\begin{equation*}
|\partial _x S_0(t,x)| \le \frac{t}{1+g(t,x)}\, \frac{2V(x)}{|1-x|}
\, + \, \frac{|1-x|}{t}\, |g(t,x)|.
\end{equation*}
From the bound \eqref{eg} on $g(t,x)$ we then conclude that both 
$S_0(t,x)$ and $\partial_x S_0(t,x)$ vanish as $t\downarrow 0$
(uniformly for $x$ in compact sets). 

Let us now consider the second derivative of $S_0(t,x)$. By
differentiating the identity \eqref{defe} we have
\begin{equation}
\label{de}
\partial_x e_{t,x} = - \frac{2}{\sqrt{4V(x)+e_{t,x}}}\,
\bigg[\int_x^1\!\frac{du}{[4V(u)+e_{t,x}]^{\frac 32}}\bigg]^{-1}.
\end{equation}
Plugging \eqref{de} in the explicit expression of 
$\partial_{xx}S_0(t,x)$ we obtain
\begin{eqnarray}
\label{ddS}
\partial_{xx}S_0(t,x) & = & \sgn(x-1) 
\frac{2V'(x)}{\sqrt{4V(x)+e_{t,x}}}
\nn \\ && + \, \frac{1}{4V(x)+e_{t,x}}\,
\bigg|\int_x^1\!\frac{du}{[4V(u)+e_{t,x}]^{\frac 32}}\bigg|^{-1} 
- \frac 1t.
\end{eqnarray}
We now write
\begin{equation*}
\bigg|\int_x^1\!\frac{du}{[4V(u)+e_{t,x}]^{\frac 32}}\bigg| =
\frac{|1-x|}{e_{t,x}^{\frac 32}} \, \big(1-D(t,x) \big)
\end{equation*}
with
\begin{eqnarray}
\label{D}
0 \;  \le \; D(t,x) & := & 1 - \frac{e_{t,x}^{\frac 32}}{1-x}
\int_x^1\!du \, \frac{1}{[4V(u)+e_{t,x}]^{\frac 32}}
\nn  \\ & \le & 1 
- \sqrt\frac{(1+g(t,x))^2}{(1+g(t,x))^2 + t^2c(x)^2},
\end{eqnarray}
where we used $V(u) \le c(x)^2(1-u)^2$, \eqref{eg}, and the identity
\begin{equation}
\label{int3}
\bigg|\int_x^1\!\frac{du}{[\gamma^2(1-u)^2 + \beta]^{\frac 32}}\bigg|
= \frac 1\beta \frac{|x-1|}{\sqrt{\gamma^2(1-x)^2+\beta}}, \qquad
\beta,\gamma>0.
\end{equation} 
Then \eqref{ddS} reads
\begin{eqnarray}
\label{ddS1}
&& \partial_{xx}S_0(t,x) \, = \, \sgn(x-1) 
\frac{2V'(x)\, t}{\sqrt{4V(x)t^2+(1-x)^2[1+g(t,x)]^2}}
\nn \\ && \qquad\qquad\quad
+ \, \frac 1t \, \bigg\{ \frac{(1-x)^2}{4V(x)t^2+(1-x)^2[1+g(t,x)]^2}\,
\frac{[1+g(t,x)]^3}{1-D(t,x)} \, - \, 1 \bigg\}. \qquad\quad
\end{eqnarray}
From the above expression and the bounds \eqref{eg} and \eqref{D}
we have $\partial_{xx}S_0(t,x) \to 0$ as $t\downarrow 0$ 
(uniformly for $x$ in compacts). 

To prove the bound \eqref{bddS} we notice that the first term on 
the r.h.s.\ of \eqref{ddS1} is bounded by $2|x|$. 
Simple algebraic manipulations yield that the second term can
be rewritten as
\begin{equation*}
\frac 1t \frac{[D(t,x)+g(t,x)]\,[1+g(t,x)]^2}
{(1-D(t,x))\, [(1+x)^2 t^2 + [1+g(t,x)]^2]} - \frac{(1+x)^2t}
{(1+x)^2 t^2 + [1+g(t,x)]^2}. 
\end{equation*}
We analyze separately the two terms above. For the second one, 
by using the bound \eqref{eg} it is easy to show that 
\begin{equation*}
\sup_{(t,x)\in\bb R_+\times\bb R} \frac{1}{c(x)} \,
\frac{(1+x)^2t} {(1+x)^2 t^2 + [1+g(t,x)]^2} < + \infty.
\end{equation*}
For the first one we first notice that, by \eqref{eg}, \eqref{D}, 
and simple computations we have
\begin{equation*}
0 \le \frac{g(t,x)+D(t,x)}{c(x)t}\le \frac{1}{c(x)t}
\bigg\{ \frac{c(x)t}{\varsh[c(x)t]} - 1 + \varth^2[c(x)t]\bigg\},
\end{equation*}
which is bounded for $(t,x)\in\bb R_+\times\bb R$. To conclude
it remains to show that 
\begin{equation*}
\sup_{(t,x)\in\bb R_+\times\bb R} 
\frac{[1+g(t,x)]^2}{(1-D(t,x))\, [(1+x)^2 t^2 + [1+g(t,x)]^2]}
< + \infty,
\end{equation*}
which can be easily checked using again \eqref{D} and \eqref{eg}. 
\qed 
      
   \begin{lemma}
   \label{t:32}
Let $e_{t,x}$ be the solution to \eqref{defe} and set 
$E_{t,x}:=e_{\ell-t,x}$, $(t,x)\in [-\ell,\ell)\times\bb R$. 
Then
   \begin{equation}
   \label{bm1}
 -\ell\le t\le\ell, \; -\infty < x\le y \le 1
\; \Longrightarrow \; E_{t,x} \ge  E_{t,y}.
   \end{equation}
Moreover, 
\begin{equation}
\label{E>}
(t,x) \in [-\ell,\ell]\times (-\infty,0] \; \Longrightarrow  \; 
\sqrt{E_{t,x}}\, \ge \, \frac{4\, e^{-(\ell-t)}}
{1+ e^{\ell-t}\, [1+x]_+}.
\end{equation}
Finally,
\begin{eqnarray}
\label{stime>}
&& \!\!\!\!\!\!\!\!
 \frac{4V(x)}{\varsh^2[(1+x)(\ell-t)]} \,\le\, E_{t,x} \,\le\,
\frac{4(x-1)^2}{\varsh^2[2(\ell-t)]},  \quad
(t,x) \in (-\ell,\ell) \times 
[1,+\infty), \qquad\;
\\ &&  \label{stime<}
\!\!\!\!\!\!\!\!
 \frac{4(x-1)^2}{\varsh^2[2(\ell-t)]} \le \, E_{t,x} \le\,
\frac{4V(x)}{\varsh^2[(1+x)(\ell-t)]}, \;\quad 
(t,x) \in (-\ell,\ell) \times (-1,1). \qquad\; 
\end{eqnarray}
   \end{lemma}
   
\proof The inequality \eqref{bm1} follows directly from the 
definition of $E_{t,x}$. By \eqref{bm1}, to prove \eqref{E>} 
it is enough to consider $x\in [-1,0]$.
In this case, from \eqref{defe} we get
\begin{eqnarray*}
\ell -t & = & 2 \int_0^1\! \frac{du}{\sqrt{4V(u) + E_{t,x}}}
- \int_{-x}^1 \! \frac{du}{\sqrt{4V(u) + E_{t,x}}}
\\ & \ge & \avarsh \frac{2}{\sqrt{E_{t,x}}} - \frac 12 \avarsh
\frac{2\,(1+x)}{\sqrt{E_{t,x}}}
\\ & \ge & \frac 12 \log \frac{16}{E_{t,x}+4\,(1+x)\sqrt{E_{t,x}}},
\end{eqnarray*}
where we used $4V(u) \le 4 (1-u)^2$ for $u\in [0,1]$ and
\eqref{int} in the second inequality, and that $\log(2y) \le 
\avarsh y \le \log(1+2y)$ for $y\ge 0$ in the last inequality. 
We thus get
$$
\sqrt{E_{t,x}} \, \ge \, \frac{8e^{-2(\ell-t)}}
{1+x + \sqrt{(1+x)^2 + 4 e^{-2(\ell-t)}}},
$$
from which the estimate \eqref{E>} follows.
Finally, to get the estimates \eqref{stime>} and \eqref{stime<} 
it is enough to insert the bounds 
\begin{equation}
\label{V><}
\left\{ \begin{array}{ll}
4(1-u)^2 \, \le\, 4V(u) \,\le\, (1+x)^2(1-u)^2 & 
\text{ if } 1\le u\le x\\
(1+x)^2(1-u)^2 \, \le\, 4V(u) \,\le\, 4(1-u)^2  
& \text{ if } x\le u\le 1
\end{array}\right.
\end{equation}
in \eqref{Etx=} and use \eqref{int}.
\qed
  
\begin{proposition}
\label{t:32b}
Let 
\begin{equation*}
\mc G_{\epsilon,L,a} := \big\{x\, :\: |\mc Z(x) | \le L,\;
d(x,\mc M) \le \epsilon^{\frac 12 -\eta},\; x(t) \ge \bam_a(t)
\,\,\,\forall\, t\in [-\ell+a,a] \big\}.
\end{equation*} 
Then, for all $\eta$ small enough there exists a real $A_5 >0$ such 
that for any $L,a>0$, and $x\in \mc G_{\epsilon, L,a}$ 
we have
\begin{equation}
\label{CL}
\varlimsup_{\epsilon\to 0} \Big|\int_{-\ell}^\ell\!dt\,
\partial_{xx} S_0(\ell-t,x(t)) + \log\ell\Big| \le A_5 \, (L+a).
\end{equation}
\end{proposition}

\proof In the sequel we shall assume that $\epsilon$ is so small
that $\epsilon^{\frac 12 -\eta}\le \frac 12$. We shall denote by $C$ 
a generic positive constant independent on $\epsilon,L,a$ whose
numerical value may change from  line to line. Fix $x\in\mc 
G_{\epsilon, L,a}$ and let $z$, $|z|\le L$, be such that 
$|x(t)-\bam_z(t)| \le \epsilon^{\frac 12-\eta}$ for any 
$t\in[-\ell,\ell]$. Setting $z^* = z + \varth(1/2)$, by the 
assumptions on $\epsilon$ we have 
\begin{equation}
\label{xt} 
\left\{
\begin{array}{ll}
\bam_a(t) \le x(t) \le \frac 12 & \text{ if } t \in [-\ell+a,a\wedge z],
\\ - \frac 12 \le x(t) \le \frac 32 & \text{ if } t\in[a\wedge z, \ell],
\\ 0 \le x(t) \le \frac 32 & \text{ if } t\in[z^*, \ell].
\end{array}\right.
\end{equation}
By \eqref{ddS}, noticing $E_{t,x} = e_{\ell-t,x}$ and
$\int_{-\ell}^{\ell-1}\!dt\, \frac{1}{\ell-t} = \log(2\ell)$,
we decompose
\begin{equation}
\label{iddS}
\int_{-\ell}^\ell\!dt\, \partial_{xx} S_0(\ell-t,x(t)) + \log\ell
= \sum_{i=1}^5 I_i  - \log 2,
\end{equation}
where
\begin{eqnarray*}
I_1 & = &  \int_{-\ell}^{-\ell+2a}\!dt\, \Big[
\partial_{xx}  
S_0(\ell-t,x(t)) + \frac{1}{\ell-t}\Big] + \int_{\ell-1}^\ell\!dt\, 
\partial_{xx} S_0(\ell-t,x(t)), \\
I_2 & = &  \int_{-\ell+2a}^{\ell-1}\!dt\, \sgn(x(t)-1) 
\frac{2V'(x(t))}{\sqrt{4V(x(t))+E_{t,x(t)}}},
\\ I_3 & = & \int_{-\ell+2a}^{a\wedge z}\!dt\, G(\ell-t,x(t)), 
\\ I_4 & = & \int_{a\wedge z}^{z^*} \!dt\, G(\ell-t,x(t)),
\\ I_5 & = &\int_{z^*}^{\ell-1}\!dt\, G(\ell-t,x(t)),
\end{eqnarray*}
with 
\begin{equation}
\label{Gtx}
G(t,x) := \frac{1}{4V(x)+e_{t,x}}\,
\bigg|\int_x^1\!\frac{du}{[4V(u)+e_{t,x}]^{\frac 32}}\bigg|^{-1}.
\end{equation}
Since $|x(t)|\le \frac 32$, by Theorem~\ref{t:31} we get
$|I_1| + |I_4| \le C\, a$. We next estimate the other 
integrals separately.

\smallskip\noindent{\it Bound on $I_2$}. Since
\begin{equation*}
\sgn(x-1) \frac{2V'(x)}{\sqrt{4V(x)}} = 2x \qquad \forall\, x\ge-1
\end{equation*}
and recalling that  $x(t)>-1$ for any $t\in [-\ell+2a,\ell]$ we have
\begin{eqnarray*}
|I_2| & \le & 2\, \int_{-\ell+2a}^{\ell-1}\!dt\, |V'(x(t))| \, \bigg[ 
\frac{1}{\sqrt{4V(x(t))}} -
\frac{1}{\sqrt{4V(x(t))+ E_{t,x(t)}}}\bigg]\\
&& + \, 2\,\int_{-\ell+2a}^{\ell-1}\!dt\, \big|x(t)-\bam_z(t)\big| 
\, + \, 2\, \bigg|\int_{-\ell+2a}^\ell\!dt\, \bam_z(t)\bigg|
\\ & \le & \int_{-\ell+2a}^{\ell-1}\!dt\, 
\frac{|V'(x(t))|}{[4V(x(t))]^{\frac 32}} \,E_{t,x(t)} 
\, + \, 4\, \epsilon^{\frac 12 -\eta}\,(\ell-a) 
+ 4 \, |a-z| \\ & \le & \frac 32\, \int_{-\ell+2a}^{\ell-1}\!dt\, 
\frac{E_{t,x(t)}}{4V(x(t))} \, + \, C\,(L+a),
\end{eqnarray*} 
where we used $|x(t)|\le \frac 32$ in the last inequality. Now, by
\eqref{bm1}, \eqref{stime>}, and \eqref{stime<} we have
\begin{eqnarray*}
&& \int_{-\ell+2a}^{\ell-1}\!dt\, 
\frac{E_{t,x(t)}}{4V(x(t))} \, \le \,   
\int_{-\ell+2a}^{\ell-1} \frac{dt}{\varsh^2[2(\ell-t)]}\,
\id_{x(t)\ge 1} \\ && \qquad\qquad + \, \int_{a\wedge z}^{\ell-1}
\frac{dt}{\varsh^2[(1+x(t))(\ell-t)]} \, \id_{x(t)< 1}
\, + \, \int_{-\ell+2a}^{a\wedge z} \!dt\,
\frac{E_{t,\bam_a(t)}}{4V(x(t))}\, \id_{x(t)< 1}
\\&&\quad  \le \,
\int_{-\ell+2a}^{\ell-1} \frac{dt}{\varsh^2[2(\ell-t)]}
\, +\, \int_{a\wedge z}^{\ell-1} \frac{dt}{\varsh^2\big[
\frac 12 (\ell-t)\big]}
+ \, 4 \, \int_{-\ell+2a}^{a\wedge z} \!dt\,
\frac{E_{t,\bam_a(t)}}{[1+\bam_a(t)]^2},
\end{eqnarray*}
where we used \eqref{xt}. The first two integrals on 
the r.h.s.\ above are readily seen to be uniformly bounded in $\ell$. For
the last one we need an upper bound for $E_{t,\bam_a(t)}$. To this end
we observe that, for $t\le a$,
\begin{eqnarray*}
\ell - t & = & \int_0^1\, \frac{du}{\sqrt{4V(u)+E_{t,\bam_a(t)}}} \,+\,
\int_0^{|\bam_a(t)|} \, \frac{du}{\sqrt{4V(u)+E_{t,\bam_a(t)}}} 
\\ & \le & \int_0^1\, \frac{du}{\sqrt{4V(u)+E_{t,\bam_a(t)}}} \,+\,
\avarth\big|\bam_a(t)\big| \\ & = & 
\int_0^1\, \frac{du}{\sqrt{4V(u)+E_{t,\bam_a(t)}}} + a - t,
\end{eqnarray*}
from which, by \eqref{El}, we get $E_{t,\bam_a(t)} \le 
C e^{-4(\ell-a)}$ for any $t\in [-\ell,a]$. Then, 
\begin{equation}
\label{em}
\int_{-\ell+2a}^{a\wedge z} \!dt\,
\frac{E_{t,\bam_a(t)}}{[1+\bam_a(t)]^2}\,
\le \, C \int_{-\ell+2a}^{a\wedge z} \!dt\, e^{-4(\ell-a)} e^{-4(t-a)}
\, \le \, C,
\end{equation}
so that $|I_2|\le C(L+a)$. 

\smallskip\noindent{\it Bounds on $I_3$ and $I_5$}.
From \eqref{V><} and \eqref{int3}, we have
 \begin{eqnarray}
\label{d>1}
G(\ell-t,x) & \le & 
\frac {E_{t,x}}{x-1} \, \frac 1{\sqrt{4V(x)+E_{t,x}}},
 \qquad \quad\, \text{ if }\quad x\in (1,+\infty),
\\ \label{d<1}
G(\ell-t,x) & \le & \frac {E_{t,x}}{1-x} \,
\frac{\sqrt{4(1-x)^2+E_{t,x}}}{4V(x)+E_{t,x}}  \quad\quad\,\,
\text{ if }\quad x\in (-1,1).
\end{eqnarray}
By \eqref{bm1}, \eqref{xt}, and \eqref{d<1},
\begin{equation*}
I_3 \le C \, \int_{-\ell+2a}^{a\wedge z} \!dt\,
\frac{E_{t,\bam_a(t)}}{[1+\bam_a(t)]^2},
\end{equation*}
and the integral on the r.h.s.\ has been bounded in \eqref{em}.
For $I_5$, we observe that, by \eqref{stime>}, \eqref{stime<}, 
\eqref{d>1}, and \eqref{d<1},
\begin{eqnarray*}
G(\ell-t,x) & \le & \frac{4}{\varsh^2[2(\ell-t)]}\, 
\frac{\varth[(1+x)(\ell-t)]}{(1+x)} \qquad  \qquad \qquad \quad
\text{ if }\quad x\in (1,+\infty),
\\ G(\ell-t,x) & \le & \frac{1}{\varsh^2[(1+x)(\ell-t)]}\,
\sqrt{4+\frac{(1+x)^2}{\varsh^2[(1+x)(\ell-t)]}} \,\quad
\text{ if }\quad x\in (-1,1).
\end{eqnarray*}
Then, recalling also \eqref{xt},
\begin{eqnarray*}
I_5 & \le & \int_{z^*}^{\ell-1} \!dt\, 
\frac{4(\ell-t)}{\varsh^2[2(\ell-t)]} \,\id_{x(t)\ge 1}
\\ && + \, \int_{z^*}^{\ell-1} \!dt\,
\frac{1}{\varsh^2[(\ell-t)]} \, \sqrt{4+\frac{1}{(\ell-t)^2}}  
\, \id_{x(t)< 1},
\end{eqnarray*}
which is uniformly bounded.
\qed

\bigskip
\noindent
\emph{Proof of (\ref{EE})}. 
We assume $\epsilon$ so small that 
\begin{equation}
\label{cond}
\epsilon^{\frac 12 -\eta} < \frac 12 e^{-2\sigma\ell}, \qquad
\ell > \frac 52, \qquad \log\ell < \sigma\ell.
\end{equation}

In particular, since $\bam_0(s) > 1 - 2e^{-2s}$ for any $s\ge 0$, 
setting $T=\frac 12 \log\ell$, for $t\in[\tau_-,\tau_+]$ we have 
$$
\xi_t = \bam_z(t) + v_t \begin{cases} < 0 & \text{ if } t - z < - T \\
> 0 & \text{ if } t-z>T \\ 
\end{cases}
$$
We shorthand $E_t=E_{t,\bam_z(t) + v_t}$ and analyze separately three cases:

\noindent $i)$ 
Assume $t\in [z-T,z+T] \cap [\tau_-,\tau_+]$. By \eqref{Etx=} 
with $x=\xi_t=\bam_z(t)+v_t$ we have:
$$
\ell \le z + T + \int_0^1\! \frac{du}{\sqrt{4V(u) + 
E_t}} + \int_0^{|\xi_t|}\! \frac{du}{\sqrt{4V(u) + E_t}}.
$$
By \eqref{cond} we have $|\xi_t| \le \bam_0(T) + \epsilon^{\frac 12
-\eta} \le  1 - \frac 12 e^{-2T}$, so that
$$
\int_0^{|\xi_t|}\! \frac{du}{\sqrt{4V(u) + E_t}} \le
\avarth\Big[ 1 - \frac 12 e^{-2T}\Big] \le T + \frac 12 \log 4,
$$
whence
\begin{equation}
\label{e1}
\ell - z - \frac 12 \log 4 - \log\ell \le  
\int_0^1\! \frac{du}{\sqrt{4V(u) + E_t}}.
\end{equation}

\noindent $ii)$ 
Let $t\in [z-\sigma\ell, z-T] \cap [\tau_-,\tau_+]$. Since $\xi_t<0$,
by \eqref{Etx=} we have:
$$
\ell \le \int_0^1\! \frac{du}{\sqrt{4V(u) + E_t}} +
t + \int_0^{|\xi_t|}\! \frac{du}{\sqrt{4V(u) + E_t}}. 
$$
By \eqref{cond} $|\xi_t| \le \bam_0(|t-z|) + \epsilon^{\frac 12 -\eta}
\le 1 - \frac 12 e^{-|t-z|} $, so that 
$$
t + \int_0^{|\xi_t|}\! \frac{du}{\sqrt{4V(u) + E_t}} 
\le t + \avarth\Big[ 1 - \frac 12 e^{-2|t-z|}\Big] \le z + \frac 12 \log 4,
$$
whence
\begin{equation}
\label{e2}
\ell - z - \frac 12 \log 4 \le  
\int_0^1\! \frac{du}{\sqrt{4V(u) + E_t}}.
\end{equation}

\noindent $iii)$ 
Finally let $t\in [z+T,\sigma\ell+z] \cap [\tau_-,\tau_+]$. Since 
$\xi_t \ge \bam_z(t) - \epsilon^{\frac 12 -\eta} \ge 
1 - 3 e^{-2(t-z)}$, by \eqref{Etx=} we have:
\begin{eqnarray*}
\ell - t & \le & \int_{1- 3 e^{-2(t-z)}}^1\! 
\frac{du}{\sqrt{4V(u) + E_t}}
\\ & \le & 
\frac{1}{2-3 e^{-2(t-z)}} 
\avarsh \frac{3 e^{-2(t-z)}\big(2-3 e^{-2(t-z)}\big)}{\sqrt{E_t}}.
\end{eqnarray*}
Recalling $t-z > \frac 12 \log\ell$, for a suitable constant $C>0$ 
and any $\epsilon$ small enough we get 
\begin{equation}
\label{e3}
E_t \le C \exp\{-4(t-z) - 4(\ell-t) + 6(\ell-t)e^{-2(t-z)} \}
\le C \exp\{-4(\ell-z) + 6\}. 
\end{equation}

By comparing \eqref{e1} and \eqref{e2} with \eqref{El} we conclude that
\begin{equation}
\label{e4}
\varlimsup_{\epsilon \to 0} 
\sup_{t\in [z-T,z+T] \cap [\tau_-,\tau_+]} 
e^{4(\ell-z)}  \ell^{-4}  E_t <\infty.
\end{equation}
The claim \eqref{EE} now follows from \eqref{e3} and \eqref{e4}.
\qed 

\section*{Acknowledgments}

\nopagebreak
The motivation of the present paper lies on comments by E.\ Presutti
on \cite{bbbdy}. We are in debt with A.\ Garroni who explained us 
the proof of Theorem~\ref{barfell}. We thank A.\ Teta for useful
discussions on the semiclassical limit. L.B.\ and P.B.\ acknowledge 
the partial support of COFIN-MIUR. S.B. aknowledges the hospitality 
at the Mathematics Department of the University of Rome `La Sapienza'.


\begin{thebibliography}{99}

\bibitem{bbbdy} L. Bertini, S. Brassesco, P. Butt\`a:
{\em Soft and hard wall in a stochastic reaction diffusion equation}
Preprint 2006.

\bibitem{Braides} A. Braides: 
{\em $\Gamma$-convergence for beginners.}
Oxford: Oxford University Press 2002.

\bibitem{CMR} M. Cassandro, I. Merola, U. Rozikov:
{\em Phase cohexistence in one dimensional Ising models with long 
range interactions.}Ê
In preparation.

\bibitem{COP} M. Cassandro, E. Orlandi, E. Presutti:
{\em Interfaces and typical Gibbs configurations for one-dimensional 
Kac potentials.}
Probab. Theory Related Fields \textbf{96}, 57--96 (1993).

\bibitem{DZ} A. Dembo, O. Zeitouni: 
{\em Large deviations techniques and applications.}
Second edition. New York: Springer 1998.

\bibitem{D} R.L. Dobrushin:
{\em Investigation of Gibbsian states for three-dimensional
lattice systems.}
Theor. Probability Appl. \textbf{18}, 253--271 (1973).

\bibitem{DS} R.L. Dobrushin, S.B. Shlosman: 
{\em The problem of translation invariance of Gibbs states at 
low temperatures}.
Soviet Scientific Reviews, Section C: Mathematical 
Physics Reviews \textbf{5}, 53--195 (1985).

\bibitem{FV} M.I.Freidlin, A.D.Wentzell: 
{\em Random perturbations of dynamical systems}. 
Second edition. New York:Springer 1998. 

\bibitem{Funaki} T. Funaki:
{\em Stochastic interface models. Lectures on probability theory 
and statistics.}
Lecture Notes in Math. \textbf{1869}, 103--274.
Berlin: Springer 2005.

\bibitem{GRS} F. Guerra, L. Rosen, B. Simon
{\em The $P(\phi)\sb{2}$ Euclidean quantum field theory as classical statistical mechanics. I.} 
Ann. of Math. \textbf{101}, 111--189 (1975).

\bibitem{HP} U.G. Haussmann, \'E. Pardoux:
{\em Time reversal of diffusions.}
Ann. Probab. \textbf{14}, 1188--1205 (1986).

\bibitem{JMS} G. Jona-Lasinio, F. Martinelli, E. Scoppola:
{\em New approach to the semiclassical limit of quantum mechanics. 
I. Multiple tunnelings in one dimension.} 
Comm. Math. Phys. \textbf{80}, 223--254 (1981).

\bibitem{KS} I. Karatzas, S.E. Shreve: 
{\em Brownian motion and stochastic calculus}. 
Second edition. New York: Springer 1991.

\bibitem{Presutti} E. Presutti:
{\em From statistical mechanics towards continuum mechanics.}
Notes of lectures given at Max Planck Institute of Leipzig (1999).

\bibitem{Robert} D. Robert
{\em Autour de l'approximation semi-classique.} 
Boston: Birkh\"auser 1987. 

\bibitem{S} H. Spohn 
{\em Large scale dynamics of interacting particles.}
Berlin: Springer 1991.

\bibitem {ST} R.H. Schonmann, N.I. Tanaka:
{\em One-dimensional caricature of phase transition.}
J. Statist. Phys. \textbf{61}, 241--252 (1990).

\bibitem{Simon} B. Simon: 
{\em Functional integration and quantum physics. }
Academic Press: New York-London 1979.

\end{thebibliography}
\end{document}